\documentclass[prd,nofootinbib,showpacs,eqsecnum,floats,draft]{revtex4}
\usepackage{amssymb}

\begin{document}
%\twocolumn[\hsize\textwidth\columnwidth\hsize\csname
%          @twocolumnfalse\endcsname
\title{Gravitational waveforms from a point particle orbiting 
a Schwarzschild black hole}
\author{Karl Martel}
\affiliation{Department of Physics, University of Guelph, Guelph,
         Ontario, Canada, N1G 2W1}
\date{\today}

\begin{abstract}
We numerically solve the inhomogeneous Zerilli-Moncrief and 
Regge-Wheeler equations in the time domain.  We 
obtain the gravitational waveforms produced by a 
point-particle of mass $\mu$ traveling around a Schwarzschild black hole 
of mass $M$ on arbitrary bound and unbound orbits.  
Fluxes of energy and angular momentum at 
infinity and the event horizon are also calculated.
Results for circular orbits, selected cases of eccentric 
orbits, and parabolic orbits are presented. 
The numerical results from the time-domain code indicate that,   
for all three types of orbital motion, black hole absorption 
contributes less than $1\%$ of the total flux, so long as the orbital radius $r_{p}(t)$ 
satisfies $r_{p}(t)>5M$ at all times.
\end{abstract}
\maketitle

%%%%%%%%%%%%%%%%%%%%%%%%%%%%%%%%%%%%%%%%%%%%%%%%%%%%%%%%%%%%%%%%%%%%%%%%%%%%
%%%%%%%%%%%%%%%%%%%%%%%%%%%%%%%%%%%%%%%%%%%%%%%%%%%%%%%%%%%%%%%%%%%%%%%%%%%%

\section{Introduction} \label{sec:intro}
Tightly bound binary systems consisting of a compact object of a few 
solar masses and a supermassive black hole of $10^{6}-10^{9}\ M_{\odot}$ are  
very promising sources of gravitational waves 
for space-based detectors such as LISA~\cite{Danzmann}.  
There is now strong evidence that most galaxies harbour a 
$10^{6}-10^{9}\ M_{\odot}$ supermassive black hole 
in their centre~\cite{Kormendy}, and that they 
are likely surrounded by a large population of 
solar-mass compact objects that reside in the galactic cusp~\cite{Gould}. 

The motion of objects in the galactic cusp is governed by 
the gravity of the supermassive black hole, but they are also constantly 
scattered due to the presence of multiple compact objects.  
For a given compact object, 
this process occurs until it settles on a highly eccentric 
orbit that is tightly bound to the central black hole.  
On such an orbit, the object passes very close to the black hole at periastron 
and it emits a significant 
amount of gravitational waves.  Capture occurs 
for those orbits that are sufficiently eccentric and have a sufficiently small periastron (on the order of $M$)~\cite{SRees}.   
In these cases, orbital evolution 
is driven by emission of gravitational waves, and the binary strongly radiates  
gravitational radiation, until the final plunge of the compact object into the central 
black hole.

The question is then to determine the rate at which solar-mass compact objects 
are captured by the central black hole and how quickly the orbits 
decay by emission of gravitational waves.  Because capture occurs when the time to evolve due to emission 
of gravitational waves is much smaller than the time to evolve due to 
diffusion and scattering, determination of the type of orbits for which capture occurs and estimate 
of capture rates are sensitive to the strength of gravitational wave emission.
Current estimates of orbital parameters for which capture occurs and associated capture 
rates are 
based on the quadrupole approximation for the emission of gravitational waves~\cite{Sigurdsson}.
Although this is well justified for large periastron,  it is not a good approximation for 
highly eccentric orbits with small periastron, those of interest for gravitational-wave astronomy.

In this paper, we consider a situation in which 
the compact object has already been 
captured by a spherically symmetric central black hole, and calculate 
the correct, general relativistic, 
rates at which the system loses energy and angular 
momentum to gravitational waves.  
We consider three types of orbits: circular, eccentric and parabolic orbits.
These calculations will then 
be used to refine capture rates estimate, 
but this will be left for future work.

At this level of approximation, the internal dynamics of the small 
compact object are irrelevant.  We treat it as a point-particle and  
base our calculations on first-order perturbations of a  
Schwarzschild black hole; this is appropriate in view of the 
small mass ratio involved. 
The gravitational waveforms produced by the orbital motion are obtained by solving 
the even parity Zerilli-Moncrief~\cite{Zerilli,Moncrief} (ZM) and the odd parity 
Regge-Wheeler~\cite{RW} (RW) equations. We work with Schwarzschild coordinates, 
for which both wave equations take the form
\begin{equation} \label{eqn:wave}
\left[-\frac{\partial^2}{\partial t^2}+\frac{\partial^2}{\partial r^{*\,2}}-V_{l}(r)\right]\psi_{lm}(r,t)=S_{lm}(r,t),
\end{equation}
where $r^{*}=r+2M\log(r/2M-1)$ is the usual tortoise coordinate, and $V_{l}(r)$ is a potential defined in Eq.~(\ref{eqn:pot}) for both modes. 
Explicit definitions for the ZM and the RW functions 
are given in Eqs.~(\ref{eqn:ZMfun}) and (\ref{eqn:RWfun}) 
of Appendix~\ref{app:pt}. 
The source term $S_{lm}(r,t)$ is of the form
\begin{equation} \label{eqn:src}
S_{lm}(r,t)=G(r,t)\delta[r-r_{p}(t)]+F(r,t)\delta '[r-r_p(t)],
\end{equation}
where a prime denotes an $r$-derivative, $r_p(t)$ denotes 
the radial position of the particle as 
a function of time, and $G(r,t)$ and $F(r,t)$ are known functions of 
$r$ and $t$ once the orbital motion of the particle is specified; 
they are given 
by Eq.~(\ref{eqn:SZM}) for the Zerilli-Moncrief equation, 
and by Eq.~(\ref{eqn:SRW}) for the Regge-Wheeler equation.

Instead of Fourier decomposing Eq.~(\ref{eqn:wave}) and 
solving in the frequency domain, 
we choose to integrate them in the time domain.
The numerical method we use was first 
developed by C.O.~Lousto and R.H.~Price~\cite{LP}, and later  
corrected by K.~Martel and E.~Poisson to yield 
second-order convergence~\cite{Martel};
it is a finite-difference scheme, based 
on the null-cones of the Schwarzschild spacetime, which 
incorporates the source term without approximating 
$\delta[r-r_p(t)]$ and $\delta '[r-r_p(t)]$.  
This method is advantageous compared to Fourier decomposition 
because of the need, in the case of highly eccentric orbits, to sum over a very large 
number of frequencies in order to obtain accurate results~\cite{Cutler,Kennefick}.
As an added bonus, the time-domain method provides the Zerilli-Moncrief and 
Regge-Wheeler functions everywhere in the spacetime.  
For each multipole moment, information about the fluxes of energy and angular momentum at infinity and through the 
event horizon is obtained by a single numerical integration.
 
Astrophysical black holes are very likely 
to be rapidly rotating and the assumption of spherical symmetry for 
the central black hole is unrealistic.  
However, removing this assumption would require a substantial revision 
of our numerical method.
The source term for the Schwarzschild 
perturbation equations can be treated exactly because, by removing the angular dependence, 
the problem is reduced to integrating a one-dimensional partial differential equation.
A divergent source term of the form of Eq.~(\ref{eqn:src}) 
then leads to a simple jump in the field at the 
particle's position, and this can easily be handled by 
finite-difference methods.  For a rotating black hole, one is faced with the 
task of solving the inhomogeneous Teukolsky equation~\cite{Teukolsky}.  
It is well known that this equation is not separable in the time domain, because 
the eigenvalues of the angular functions are frequency dependent.  
Insisting on working in the time domain leaves a 
two-dimensional partial differential equation to integrate.  Unfortunately, 
a $\delta$-function source  
no longer leads to a simple jump at the position of the particle: 
the field is now (logarithmically) divergent at this location.
Standard finite-difference methods are inadequate to deal with 
this type of behaviour and cannot be used.  
The problem can be circumvented by smearing the particle around its position 
(for example by using narrow 
Gaussian functions instead of $\delta$-functions).  
This eliminates the divergence in the source term and, consequently, in the field; 
standard finite-difference methods can then be applied.  Such an approach 
has been used to obtain gravitational waveforms 
produced by a particle on an equatorial circular 
orbit of the Kerr black hole~\cite{Khanna},
but the error introduced by smearing the particle is difficult to ascertain.  
By specializing to Schwarzschild, comparison with the present work will allow such a determination; 
this is another important justification for the work presented here.
With this application in mind, we consider orbits with a wide range of eccentricities and 
semi-major axis,  and do not 
necessarily restrict ourselves to highly eccentric orbits.

The paper is organized as follows. 
In Sec.~\ref{sec:orbit} we describe
the orbital parametrization of bound and marginally-bound geodesics of 
the Schwarzschild spacetime.
In Sec.~\ref{sec:fluxes} we provide a relation between the Zerilli-Moncrief and Regge-Wheeler functions 
and the gravitational waveforms at infinity and near the event horizon;  
from these relations, the fluxes of energy and angular momentum 
can be calculated.  
In Sec.~\ref{sec:accuracy}, we provide a discussion of 
numerical issues that limit the accuracy with which we can determine the fluxes.  
In Sec.~\ref{sec:circular}, Sec.~\ref{sec:eccentric}, and Sec.~\ref{sec:parabolic}, 
we present our results for the gravitational waveforms and fluxes for circular, eccentric, 
and parabolic orbits, respectively.  In Sec.~\ref{sec:Conclusion}, we summarize our findings. 
Appendix~\ref{app:pt} contains a brief summary of first-order black hole perturbation theory, 
while Appendix~\ref{app:Fluxes} contains a detailed derivation of the flux formulae 
at infinity (Appendix~\ref{app:org}) and through the event horizon (Appendix~\ref{app:irg}).

%%%%%%%%%%%%%%%%%%%%%%%%%%%%%%%%%%%%%%%%%%%%%%%%%%%%%%%%%%%%%%%%%%%%%%%%%%%%
%%%%%%%%%%%%%%%%%%%%%%%%%%%%%%%%%%%%%%%%%%%%%%%%%%%%%%%%%%%%%%%%%%%%%%%%%%%%

\section{Orbital parametrization} \label{sec:orbit}
Following C.~Cutler {\it et al.}~\cite{Cutler}, we introduce $p$, 
the semi-latus rectum, and $e$, the eccentricity, as orbital parameters.  
They are defined so that the periastron and apastron are at $pM/(1+e)$ 
and $pM/(1-e)$, respectively.  In terms of these parameters, 
the energy and angular momentum per unit mass of a point particle are
\begin{eqnarray}
\tilde{E}^2&=&\frac{(p-2-2e)(p-2+2e)}{p(p-3-e^2)}, \nonumber \label{eqn:epu} \\
\tilde{L}^2&=&\frac{M^2p^2}{p-3-e^2}. \label{eqn:lpu}
\end{eqnarray}
For $e=0$ the periastron and apastron coincide, 
and the orbit is circular.  In the interval $0\leq e <1$, 
the motion occurs between two turning points, while for $e=1$, 
the apastron is pushed back to infinity and 
the motion is parabolic\footnote{In analogy with Newtonian mechanics, 
we use the term ``parabolic'' for marginally bound orbits: they have $e=1$ ($\tilde{E}=1$), but 
the trajectories traced out are {\em not} parabolae, except in the limit $p\gg 1$.}.  In all cases, stable 
orbits exist only if $p> 6+2e$. 

\begin{figure}[!htb]
\vspace*{2.5in}
\special{hscale=31 vscale=31 hoffset=-13.0 voffset=190.0
         angle=-90.0 psfile=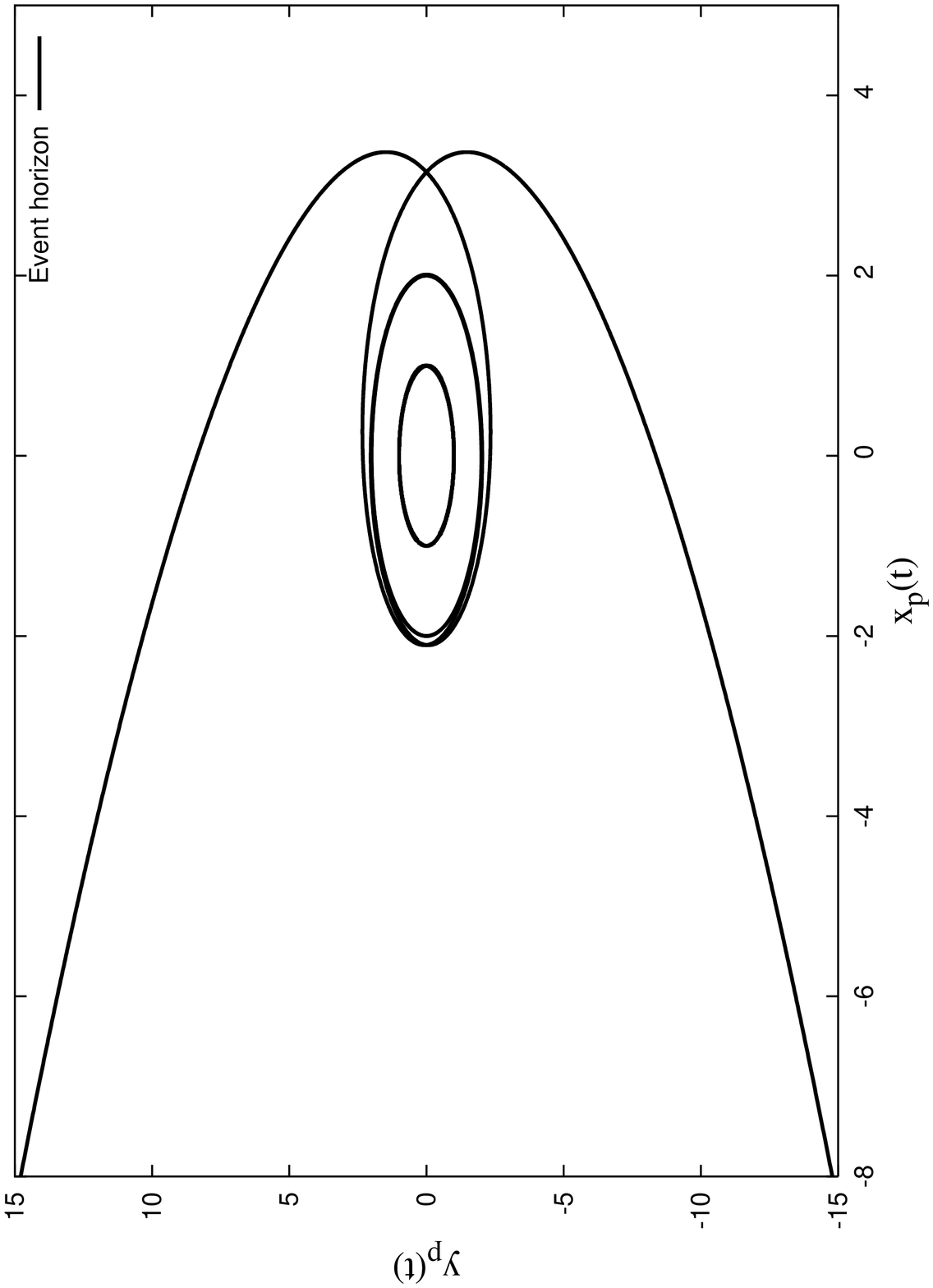}
\special{hscale=31 vscale=31 hoffset=225.0 voffset=190.0
         angle=-90.0 psfile=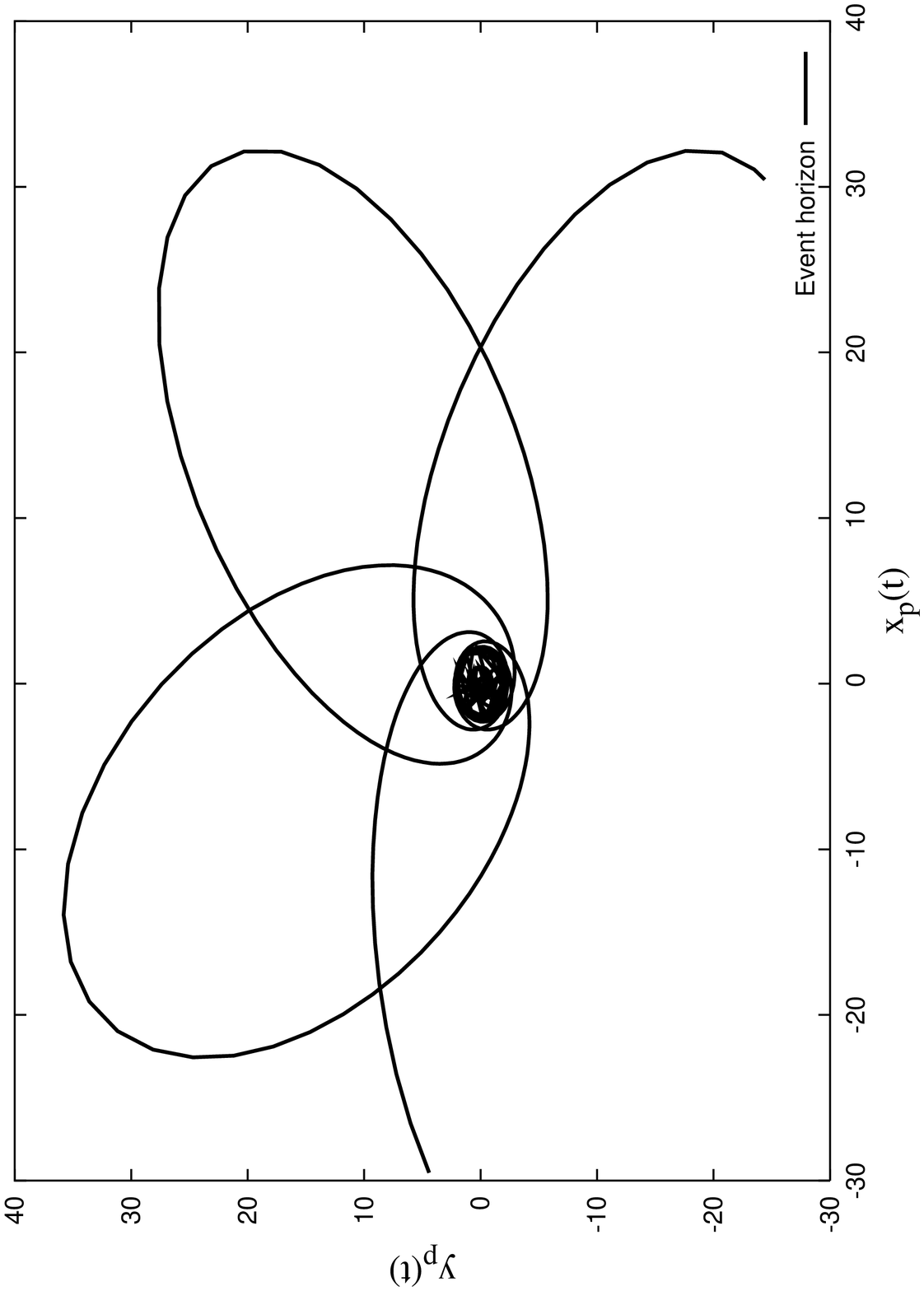}         
\caption{In the left panel, we display the trajectories in 
the $x_{p}$-$y_{p}$ plane for a geodesic with $e=1$, and $p=8.001$.
For this choice of parameters, the particle orbits the black hole 
approximately four times before leaving the central region.
In the right panel, we display a $e=0.9$ and $p=7.8001$ geodesic.  
When the particle reaches the periastron, 
it orbits the black hole on a quasi-circular orbit 
for approximately six cycles.  In both cases, the exact number of cycles 
is given by Eq.~(\ref{eqn:Ncycle}).
}\label{fig:orbits}
\end{figure}
The position of the particle at time $t$ is given by the coordinates 
$(r_p(t), \varphi_p(t), \theta_p=\pi/2)$.  
Inspired by the solution to the two-body problem in Newtonian mechanics, 
the radial position of the particle is expressed as
\begin{equation} \label{eqn:rp}
r_{p}(\chi)=\frac{pM}{1+e\cos\chi}, 
\end{equation}
where $\chi$ is a parameter along the orbit. This is 
well behaved at the turning points ($\chi=0$, $\pi$), 
which facilitates the numerical integration 
of the geodesic equations for the time and angular coordinates.  
In terms of $\chi$, these are~\cite{Cutler}
\begin{eqnarray} 
\frac{d}{d\chi} t           &=& M p^2\frac{(p-2-2e)^{1/2}(p-2+2e)^{1/2}}{(p-2-2e\cos\chi)(1+e\cos\chi)^{2}(p-6-2e\cos\chi)^{1/2}}, \label{eqn:dt}\\
\frac{d}{d\chi} \varphi_{p} &=& \frac{p^{1/2}}{(p-6-2e\cos\chi)^{1/2}}. \label{eqn:dphi}
\end{eqnarray}
The first of these equation can be numerically inverted to yield $\chi(t)$; 
knowledge of $r_p(\chi)$ and $\varphi_p(\chi)$ is then equivalent to $r_p(t)$ and $\varphi_p(t)$.

The geodesic equations given by Eqs.~(\ref{eqn:dt}) and (\ref{eqn:dphi}) are integrated using the 
Burlisch-Stoer method~\cite{numrec}, and we chose the initial conditions as follows.
The gravitational waveforms are extracted as functions of time at a location $r^{*}_{\rm obs}$.  We take the initial 
moment $t=-t_{o}<0$ to be the one at which the particle is at periastron ($\chi = 0$, $r_{p}(-t_{o})=Mp/(1+e)$, and $\varphi_{p}(-t_{o})=0$).  
We set $t_{o}$ equal to the light travel time between the periastron and the 
observation point.  Thus, radiation emitted at the initial moment will reach the observer at $t\approx 0$.

This parametrization  
of the geodesic is suitable for bound and unbound orbits of the 
Schwarzschild spacetime; for $e<1$, 
the parameter $\chi$ can take any real value, 
whereas for $e\geq 1$,  
it is confined to $-\pi/e \leq \chi \leq \pi/e$.  In this paper, we 
consider circular orbits, selected cases of eccentric orbits, 
and parabolic orbits ($e=1$), 
but the code is capable of producing gravitational waveforms for any value of $e$. 
For any $p$ and $e$, the particle orbits the central black hole a number $N=\Delta \varphi_{p}/(2\pi)$ 
of times before moving out of the central region.  Integrating Eq.~(\ref{eqn:dphi}) 
over one radial period yields~\cite{Cutler}
\begin{eqnarray} \label{eqn:Ncycle}
N&=&\frac{2}{\pi}\sqrt{\frac{p}{p-6+2e}} \ K\Bigg(\frac{4e}{p-6+2e}\Bigg), 
\end{eqnarray}
where $K(m)=\int_{0}^{\pi/2} d x(1-m \sin^2 x)^{-1/2}$ is 
the complete elliptic integral of the first kind.  
To visualize the trajectories, 
we introduce $x_{p}(t)=r_{p}(t)/(2M)\cos(\varphi_{p}(t))$ and $y_{p}(t)=r_{p}(t)/(2M)\sin(\varphi_{p}(t))$.
In Fig.~\ref{fig:orbits}, we display trajectories in the $x_{p}$-$y_{p}$ plane 
for $p=7.8001$ and $e=0.9$ (left), and $p=8.001$ and $e=1$ (right).  
In both cases, the number of times the particle orbits the central black hole is large.  
This is because $p$ is close to the critical value $6+2e$ at which $N$ diverges.  
In these cases, gravitational-wave emission is dominated by the quasi-circular 
portion of the orbit near periastron.  The total energy emitted is 
then well approximated by $E=NE_{circular}$, 
where $N$ is the (divergent) number of orbits, $P$ the period of a 
circular orbit at $r_{p}=Mp/(1+e)$, and $E_{circular}$ is the energy emitted 
by a particle on such an orbit; a similar approximation holds for $L$.

%%%%%%%%%%%%%%%%%%%%%%%%%%%%%%%%%%%%%%%%%%%%%%%%%%%%%%%%%%%%%%%%%%%%%%%%%%%%
%%%%%%%%%%%%%%%%%%%%%%%%%%%%%%%%%%%%%%%%%%%%%%%%%%%%%%%%%%%%%%%%%%%%%%%%%%%%

\section{Waveforms, energy and angular momentum radiated}
To numerically evolve Eq.~(\ref{eqn:wave}) initial conditions 
must be provided for the gravitational perturbations.  
The manner in which the initial configuration of the gravitational field 
influences the subsequent evolution has been studied previously for radial 
geodesics~\cite{Martel}.  For bound geodesics, the motion is quasi-periodic 
and waiting a sufficiently long time eliminates the contribution from the initial conditions, which simply 
propagates away.  
For marginally-bound geodesics, we chose the initial position of the particle 
to be very far from the periastron.  Far away from the black hole, the velocity of the particle 
is small and it takes much longer for the particle to reach periastron 
than for the initial gravitational-wave content to escape from the system.  
At the point where the emission of radiation is strongest, there is 
no trace left of the initial configuration of the gravitational perturbations.  
This allows us to completely avoid problems related to the choice of initial data for both 
bound and marginally-bound geodesics.  We chose zero initial conditions for the 
gravitational perturbations, acknowledging that this is inconsistent (creating 
the particle from nothing violates energy-momentum conservation), but recognizing that 
artifacts of this choice disappear in time.  Fluxes may then be computed 
reliably after waiting a sufficiently long time.

%%%%%%%%%%%%%%%%%%%%%%%%%%%%%%%%%%%%%%%%%%%%%%%%%%%%%%%%%%%%%%%%%%%%%%%%%%%%
%%%%%%%%%%%%%%%%%%%%%%%%%%%%%%%%%%%%%%%%%%%%%%%%%%%%%%%%%%%%%%%%%%%%%%%%%%%%

\subsection{Far-zone fluxes and black hole absorption} \label{sec:fluxes}
We first provide a 
short summary of the relations between the Zerilli-Moncrief 
and the Regge-Wheeler functions and the radiative portion of the metric perturbation  
at infinity and at the event horizon, as well as flux formulae used throughout the paper. 
Complete summaries of first-order black hole 
perturbation theory and flux calculations are 
relegated to Appendices~\ref{app:pt} and \ref{app:Fluxes}, respectively.
In the radiation zone, the two gravitational-wave polarizations are 
related to the Zerilli-Moncrief and the Regge-Wheeler functions by
\begin{equation} \label{eqn:hphcO}
h_{+}-\imath h_{\times}=\frac{1}{2r}\sum_{lm}\sqrt{\frac{(l+2)!}{(l-2)!}}\left(\psi_{ZM}(t)-2\imath\int^{t}dt' \psi_{RW}(t')\right) \,_{- 2}Y^{lm}(\theta,\varphi).
\end{equation}
Similarly, when $r\rightarrow 2M$, the two gravitational-wave polarizations are 
given by
\begin{equation} \label{eqn:hphcI}
h_{+}+\imath h_{\times}=\frac{1}{4M}\sum_{lm}\sqrt{\frac{(l+2)!}{(l-2)!}}\left(\psi_{ZM}(t)-2\imath\int^{t}dt' \psi_{RW}(t')\right) \,_{2}Y^{lm}(\theta,\varphi).
\end{equation}
In these equations, $_{s}Y^{lm}$ are spherical harmonics of 
spin-weight $s$~\cite{Goldberg}.

From Isaacson's stress-energy tensor for gravitational waves~\cite{Isaacson}, as well as Eqs.~(\ref{eqn:hphcO}) and (\ref{eqn:hphcI}),  
we calculate the energy flux in each multipole moment to be
\begin{eqnarray} \label{eqn:eflux}
\dot{E}^{\infty,{\rm eh}}_{lm} & = & \left \{ 
\begin{array}{lr}
\frac{1}{64\pi}\frac{(l+2)!}{(l-2)!}|\dot{\psi}_{ZM}|^2 & \ \textrm{, $l+m$ even}
\vspace{0.2cm} \\
\frac{1}{16\pi}\frac{(l+2)!}{(l-2)!}|\psi_{RW}|^2       & \textrm{, $l+m$ odd}, 
\end{array} \right.
\end{eqnarray}
and the angular momentum flux to be 
\begin{eqnarray} \label{eqn:lflux}
\dot{L}^{\infty,{\rm eh}}_{lm} & = & \left \{ 
\begin{array}{lr}
\frac{\imath m}{128\pi}\frac{(l+2)!}{(l-2)!}\ \dot{\psi}_{ZM}\tilde{\psi}_{ZM} + c.c. & \ \textrm{, $l+m$ even}
\vspace{0.2cm} \\
\frac{\imath m}{32\pi}\frac{(l+2)!}{(l-2)!}\ \psi_{RW}\int dt \tilde{\psi}_{RW} + c.c.       & \textrm{, $l+m$ odd},
\end{array} \right.
\end{eqnarray}
where a ``\,\,$\tilde{\,\,}$\,\,'' over a quantity denotes complex 
conjugation, $c.c.$ is the complex conjugate,  and the $l$ and $m$ indices 
are implicit on $\psi_{ZM}$ and $\psi_{RW}$.  In Eqs.~(\ref{eqn:eflux}) and (\ref{eqn:lflux}), 
$\dot{E}^{\infty}_{lm}$ and $\dot{L}^{\infty}_{lm}$ denote the fluxes across a surface $r={\rm const.}\ \rightarrow \infty$, 
while $\dot{E}^{{\rm eh}}_{lm}$ and $\dot{L}^{{\rm eh}}_{lm}$ denote the fluxes through a 
surface $r={\rm const.}\ \rightarrow 2M$.  
The fluxes at infinity are calculated using the Zerilli-Moncrief and 
Regge-Wheeler functions extracted at $r^{*}=r^{*}_{\rm obs}$, where $r^{*}_{\rm obs}$ is large and positive, 
while for the horizon fluxes, 
they are extracted at $r^{*}=r^{*}_{\rm eh}$, where $r^{*}_{\rm eh}$ is large 
and negative.
Once $\dot{E}_{lm}$ and $\dot{L}_{lm}$ are known, the total fluxes are obtained 
by summing over all modes:
\begin{eqnarray} \label{eqn:ELtot}
\dot{E}^{\infty,{\rm eh}}&=&\sum_{l=2}^{\infty}\dot{E}^{\infty,{\rm eh}}_{l}, \qquad \quad \dot{E}^{\infty,{\rm eh}}_{l}= \dot{E}^{\infty,{\rm eh}}_{l0} + 2\sum_{m=1}^{l}\dot{E}^{\infty,{\rm eh}}_{lm}, \label{eqn:Etot}\\
\dot{L}^{\infty,{\rm eh}}&=&\sum_{l=2}^{\infty}\dot{L}^{\infty,{\rm eh}}_{l}, \ \textrm{and} \quad \dot{L}^{\infty,{\rm eh}}_{l}= 2\sum_{m=1}^{l}\dot{L}^{\infty,{\rm eh}}_{lm}; \label{eqn:Ltot}
\end{eqnarray}
there is no $m=0$ contribution to the angular momentum flux, and the factor of 2 in front of $\dot{E}^{\infty,{\rm eh}}_{lm}$ and $\dot{L}^{\infty,{\rm eh}}_{lm}$ comes from folding the $m<0$ 
contributions over to $m>0$ (see Appendix~\ref{app:pt}).  
In a slow-motion, weak-field approximation the quadrupole moment dominates ($l=2$ and $m=2$) 
and the total energy and angular momentum radiated over one orbital period are~\cite{Peters}
\begin{eqnarray} \label{eqn:QuadAppTot}
E_{Q}(p,e)&=&\frac{64\pi}{5}\frac{\mu^2}{M}\left(1+\frac{73}{24}e^2+\frac{37}{96}e^4\right)p^{-7/2}, \\
L_{Q}(p,e)&=&\frac{64\pi}{5}\mu^2 \left(1+\frac{7}{8}e^2\right)p^{-2}.
\end{eqnarray}
The average energy and angular momentum radiated per unit time, 
defined by performing an orbital average, are 
\begin{eqnarray}\label{eqn:QuadApp}
\left<\dot{E}_{Q}\right>&=&\frac{32}{5}\left(\frac{\mu}{M}\right)^2 \frac{(1-e^2)^{3/2}}{p^{5}}\left(1+\frac{73}{24}e^2+\frac{37}{96}e^4\right), \nonumber \\
\left<\dot{L}_{Q}\right>&=&\frac{32}{5}\frac{\mu^2}{M}\frac{(1-e^2)^{3/2}}{p^{7/2}}\left(1+\frac{7}{8}e^2\right).
\end{eqnarray}
For the binaries considered in this paper, the slow-motion 
and weak-field approximations break down, and 
the fluxes must be computed using Eqs.~(\ref{eqn:Etot}) and (\ref{eqn:Ltot}).  
Numerically we cannot perform 
the infinite sums, and we truncate them at a finite value $l_{\rm max}$.
In the next subsection, we explain the criteria used to choose $l_{\rm max}$ and discuss the 
overall accuracy of the time-domain computation.

%%%%%%%%%%%%%%%%%%%%%%%%%%%%%%%%%%%%%%%%%%%%%%%%%%%%%%%%%%%%%%%%%%%%%%%%%%%%
%%%%%%%%%%%%%%%%%%%%%%%%%%%%%%%%%%%%%%%%%%%%%%%%%%%%%%%%%%%%%%%%%%%%%%%%%%%%

\subsection{Accurate determination of the fluxes: numerical issues} \label{sec:accuracy}
In order to calculate the fluxes to a relative accuracy 
$\varepsilon$ (we use $\varepsilon=0.01$), 
we need to consider three sources of error: discretization of 
Eq.~(\ref{eqn:wave}), 
effects of the finite size of our computational grid, 
as well as truncation of the sums in Eqs.~(\ref{eqn:Etot}) and (\ref{eqn:Ltot}).

Firstly, discretization of Eq.~(\ref{eqn:wave}) introduces numerical truncation errors. 
In a previous paper, we showed that the Lousto and Price algorithm 
can be corrected to yield second-order convergence~\cite{Martel}, i.e. 
truncation errors scale as $\Delta t^2$, 
with $\Delta t$ denoting the numerical stepsize for evolution.  
Throughout this work we generated gravitational waveforms by setting 
$\Delta t=0.1 (2M)$ in the numerical algorithm; 
this proved sufficient to determine the fluxes at infinity to the desired $1\%$ accuracy. 
However, for a given stepsize the fluxes through the event horizon are never determined as accurately as 
the fluxes at infinity. 
The gravitational waves flowing through the event horizon are weaker than the ones 
escaping to infinity, and, because of this difference in scales, horizon fluxes are determined with an accuracy 
$\lesssim 5 \%$.  But we will see below that horizon fluxes {\em never} amount to more than a few percents of the total fluxes. 
The lower accuracy with which black hole absorption is determined is then sufficient for our goal of $1\%$ overall accuracy.

Secondly, the expressions for the fluxes displayed in Eqs.~(\ref{eqn:eflux}) and (\ref{eqn:lflux}) hold 
only asymptotically ($r^{*} \rightarrow \pm \infty$).  Numerically we are forced to extract 
the waveforms at finite $r^{*}$ values, and this introduces finite-size effects in our results.  
Numerical efficiency requires a small computational grid, but 
accuracy requires the 
waveforms to be extracted at a large and positive $r^{*}_{\rm obs}$ and at a large and negative 
$r^{*}_{\rm eh}$.  
The flux formulae developed here are based on the stress-energy tensor 
for gravitational waves, as constructed by Isaacson~\cite{Isaacson}.  The validity 
of the construction depends on $\lambda/{\mathcal R}\ll 1$ 
being satisfied, where $\lambda$ is a wavelength of the radiation and 
${\mathcal R}$ a typical radius of curvature.  To calculate the fluxes far from the black hole, 
we extract the waveforms in an approximate radiation zone 
defined by $\lambda/r_{\rm obs}\ll 1$, 
where $\lambda^{-1}\sim (M/R_{p}^{3})^{1/2}$ and $R_{p}$ is a typical orbital radius.
The radiation zone is then defined by $R_{p}/r_{\rm obs}(R_{p}/M)^{1/2}\ll 1$.  
For relativistic motion $R_{p}\sim M$ and by 
imposing $R_{p}/r_{\rm obs}<\varepsilon$, we make an error of 
order $\varepsilon$ in approximating the radiation zone.  
This is somewhat different 
from the criteria for the validity of Isaacson's 
stress-energy tensor, but since ${\mathcal R}^{-1}\sim(M/r^{3}_{\rm obs})^{1/2}$, 
we have that $\lambda/{\mathcal R}\sim (R_{p}/r_{\rm obs})^{3/2}\sim \varepsilon^{3/2}$, 
and the use of the stress-energy tensor is justified.
In practice we also imposed $r^{*}_{\rm obs}>750(2M)$.  
At the horizon, the situation is somewhat different.  The typical 
radius of curvature is ${\mathcal R}\sim \sqrt{2}2M$, but the radiation is blueshifted 
so that $\lambda\sim f_{\rm eh}(R_{p}^3/M)^{1/2}\rightarrow 0$, 
where $f_{\rm eh}=1-2M/r_{\rm eh}$.  
The requirement $\lambda/{\mathcal R}\ll 1$ then  
translates to $f_{\rm eh} (R_{p}/(2M))^{3/2} < \varepsilon$.   We used 
$R^{*}_{p}/|r^{*}_{\rm eh}|<\varepsilon$, 
as well as $r^{*}_{\rm eh}<-750 (2M)$, which amply satisfies 
the above requirement.  This yielded good results, but 
a better, more efficient choice would 
have been $r_{\rm eh} =2M[1+(2M/R_{p})^{3/2}\varepsilon]$.  
With these choices of $r^{*}_{\rm obs}$ and $r^{*}_{\rm eh}$, we are making an 
error of {\em at most} order $\varepsilon$ in determining the fluxes at infinity and through the event horizon, respectively.

Finally, the last source of error limiting the accuracy of the determination of 
the fluxes arises from truncating the sums in 
Eqs.~(\ref{eqn:Etot}) and (\ref{eqn:Ltot}) at a finite value $l_{\rm max}$.  The error is made small enough for our requirements by demanding that 
\begin{eqnarray} \label{eqn:lmax}
\varepsilon \equiv {\rm max}\Bigg(\frac{\dot{E}^{\infty}_{l_{max}}}{\dot{E}^{\infty}}, \frac{\dot{L}^{\infty}_{l_{max}}}{\dot{L}^{\infty}}\Bigg)\leq 1 \%
\end{eqnarray}
be satisfied.
Typically, Eq.~(\ref{eqn:lmax}) is satisfied with $l_{max}$  
given by $\left(p/(1+e)\right)^{-(l_{max}-2)}<\varepsilon$, which is known to hold 
for circular orbits~\cite{Poisson}.  Note that 
because $\dot{E}^{\infty}_{l_{\rm max}}$ and $\dot{L}^{\infty}_{l_{\rm max}}$ are included in the sum, 
the error comes from neglecting terms starting at $l=l_{\rm max}+1$.  
In effect, the relative error made from neglecting these terms 
is much smaller than $1\%$.  In the sequel, we will return with empirical estimates 
of our numerical errors; these will confirm the preceding qualitative discussion.

%%%%%%%%%%%%%%%%%%%%%%%%%%%%%%%%%%%%%%%%%%%%%%%%%%%%%%%%%%%%%%%%%%%%%%%%%%%%
%%%%%%%%%%%%%%%%%%%%%%%%%%%%%%%%%%%%%%%%%%%%%%%%%%%%%%%%%%%%%%%%%%%%%%%%%%%%

\subsection{Circular orbits} \label{sec:circular}
\begin{figure}[!htb]
\vspace*{2.5in}
\special{hscale=31 vscale=31 hoffset=-13.0 voffset=190.0
         angle=-90.0 psfile=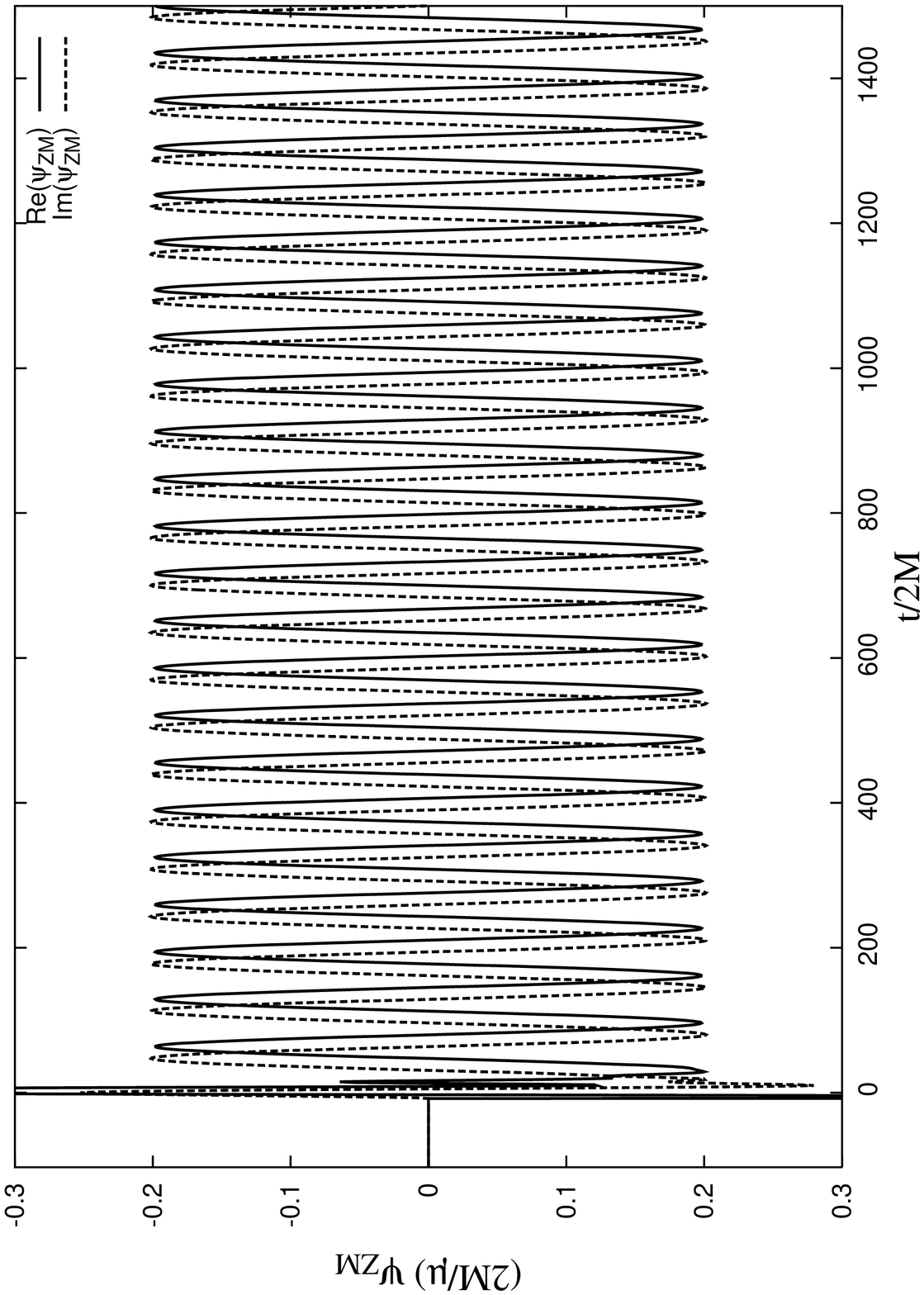}
\special{hscale=31 vscale=31 hoffset=225.0 voffset=190.0
         angle=-90.0 psfile=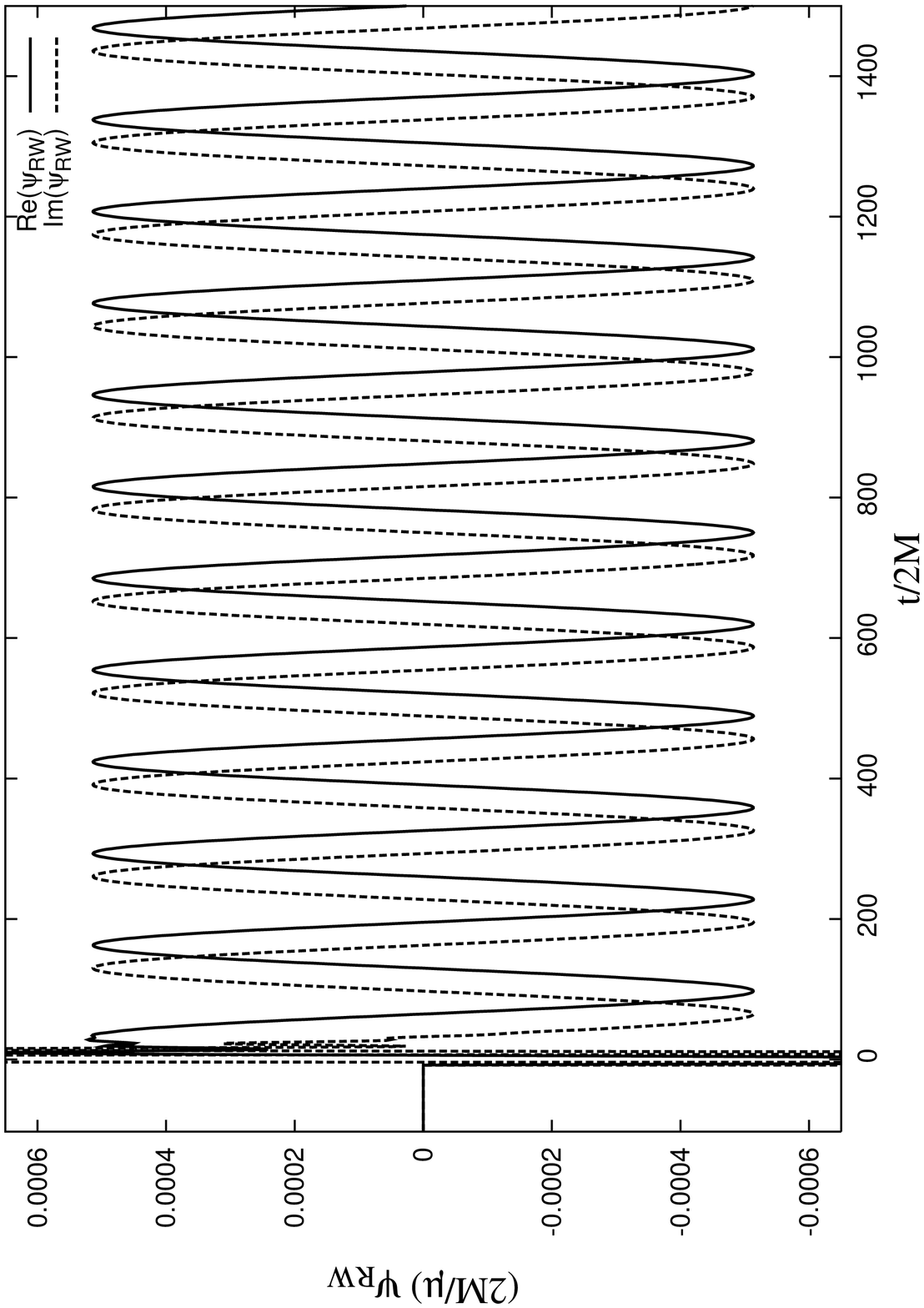}
\caption{The dominant radiation modes for the 
Zerilli-Moncrief (left, $l=2$ and $m=2$) and Regge-Wheeler 
(right, $l=2$ and $m=1$) 
functions for a particle orbiting the black hole at $r_{p}=12M$.  
At early times, the waveforms are dominated by 
the initial data content.   
We calculate the energy and angular momentum fluxes after a 
time $t/(2M)=350.0$}\label{fig:circular}
\end{figure}

For circular orbits, $e=0$ and the radius of the orbit is $r_{p}=pM$.
In Fig.~\ref{fig:circular} we display typical gravitational waveforms 
emitted by a particle traveling on a circular orbit.  Both waveforms have the same pattern: 
The field oscillates with an angular frequency given by $m \Omega$, where $\Omega=M^{-1}p^{-3/2}$ is the 
orbital angular velocity and $m$ is the multipole index. The left panel 
contains the dominant quadrupolar mode ($l=2$ and $m=2$), while the right panel 
contains the dominant odd parity mode ($l=2$ and $m=1$).

\begin{figure}[!htb]
\vspace*{2.5in}
\special{hscale=31 vscale=31 hoffset=-13.0 voffset=190.0
         angle=-90.0 psfile=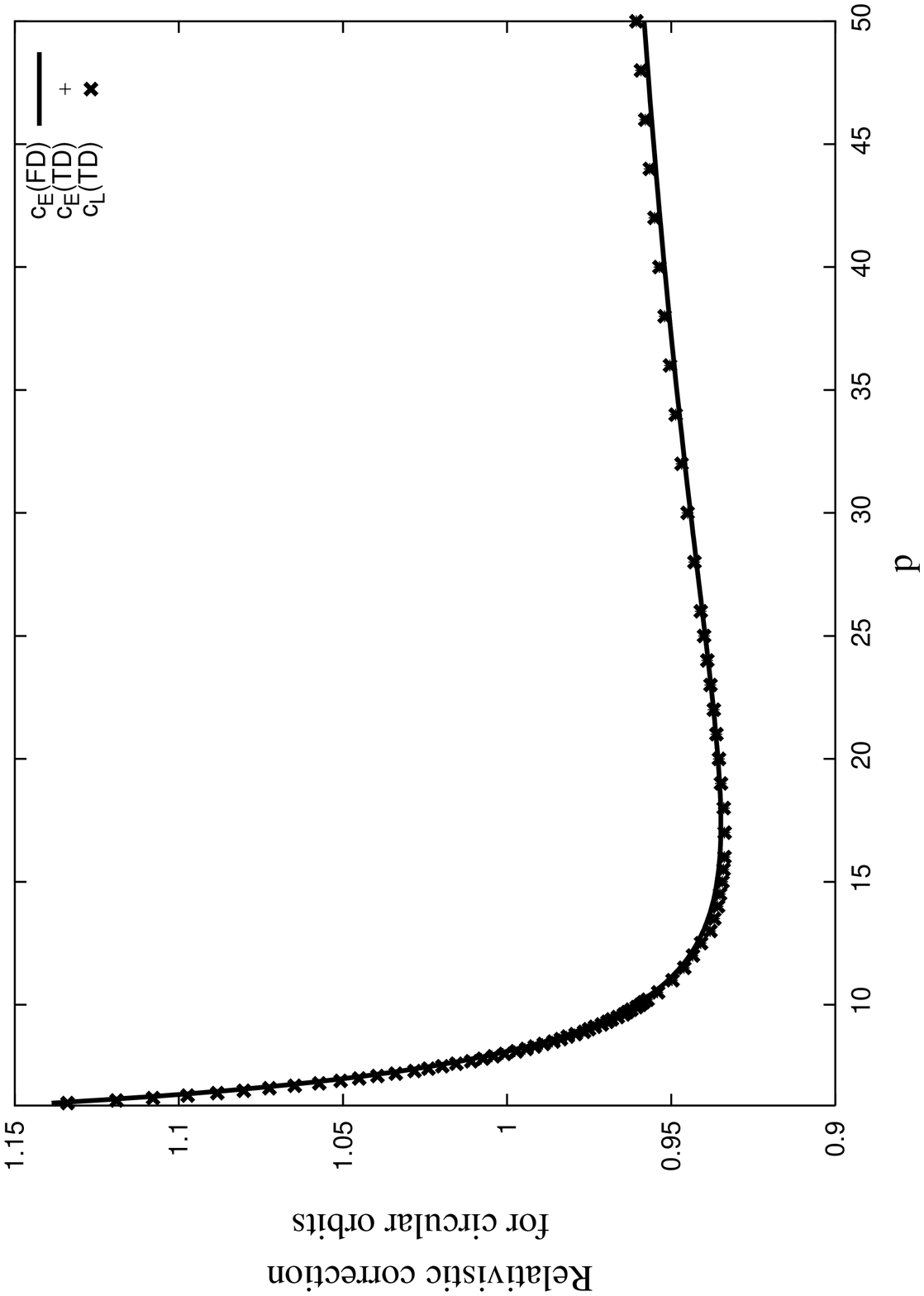}
\special{hscale=31 vscale=31 hoffset=225.0 voffset=190.0
         angle=-90.0 psfile=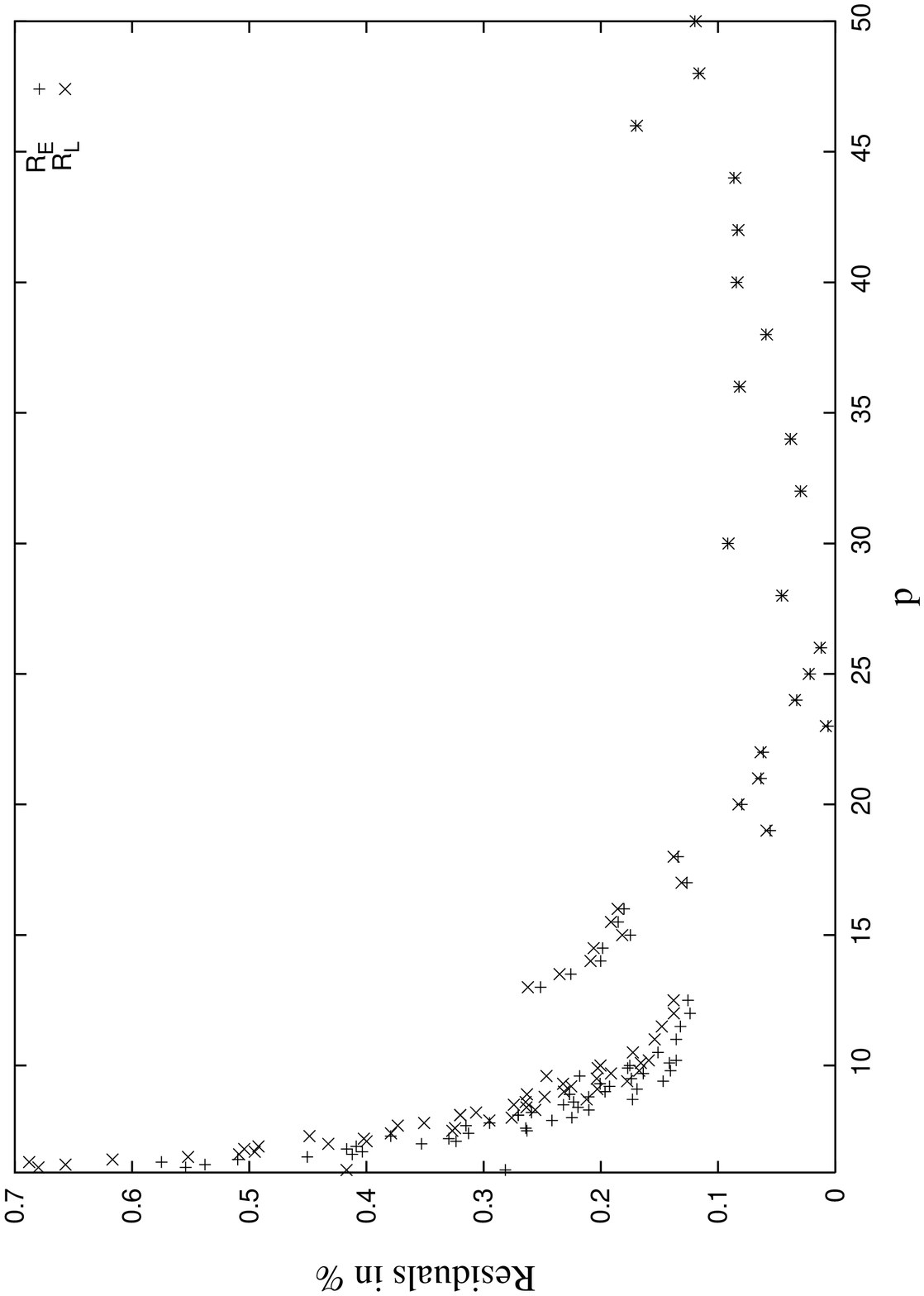}         
\caption{In the left panel, we display $c_{E}(FD)$, as well as $c_{E}(TD)$ and $c_{L}(TD)$, 
as functions of $p$.  Both $c_{E}$ and $c_{L}$ slowly approach 1 from below for large $p$.  For 
small values of $p$, the coefficients approach 1.15 as $p$ approaches 6.
In the right panel, we display the residuals 
$R_{E}$ and $R_{L}$ as defined in the text.  Using the time-domain method, the fluxes are 
calculated accurately to $0.7\%$ for $p=6.0001$, and to $0.2\%$ for large values of $p$.
}\label{fig:EL_norm}
\end{figure}
The code outputs $\dot{E}_{GR}$ and $\dot{L}_{GR}$ 
directly, but it proves convenient to express the fluxes in terms of 
$c_{E}$ and $c_{L}$: coefficients that remain close to 1 for all values of $p$.
The total fluxes are calculated using Eqs.~(\ref{eqn:Etot}) and (\ref{eqn:Ltot}) and we express the 
numerically obtained results in the form 
\begin{eqnarray}
\dot{E}^{\infty}_{GR}(p)&=& c_{E} \ \dot{E}_{Q}(p,0), \nonumber \\
\dot{L}^{\infty}_{GR}(p)&=& c_{L} \ \dot{L}_{Q}(p,0),
\end{eqnarray}
where $\dot{E}_{Q}(p,e)$ and $\dot{L}_{Q}(p,e)$ are given by Eq.~(\ref{eqn:QuadApp}) 
above with $e=0$.  For circular orbits, we should find $\dot{E}=\Omega \dot{L}$ and therefore, $c_{E}=c_{L}$.

Circular orbits have been studied extensively and we use them 
to quantitatively 
test the accuracy of the time-domain method.  We perform a 
comparison of our results with the time-domain code (TD) 
with results obtained in the frequency domain (FD) by 
E.~Poisson~\cite{Poisson_circular}.
In the left panel of Fig.~\ref{fig:EL_norm} we display $c_{E}(TD)$, 
$c_{L}(TD)$, and $c_{E}(FD)$.  In the right panel, 
we display the residuals, $R_{E}=100 |c_{E}(TD)-c_{E}(FD)|/c_{E}(FD)$ 
and $R_{L}=100 |c_{L}(TD)-c_{E}(FD)|/c_{E}(FD)$.
In the interval $6 <p\leq 50$, the time-domain code 
reproduces the frequency domain calculations to $0.7\%$ or better, 
with the best agreement occurring for large values of $p$.

\begin{table}[th!]
\caption{\label{tab:cenrg} Energy and angular momentum fluxes 
for circular orbits, calculated using a time domain (TD) code, are compared with 
fluxes calculated by E.~Poisson using a frequency domain (FD)
approach~\cite{Poisson_circular}. Here we chose $p=7.9456$ and $p=46.062$.  
The energy fluxes are in units of $\left(M/\mu\right)^2$, and the angular momentum 
fluxes are in units of $M/\mu^2$. They are 
calculated at $r^{*}_{\rm obs}=1500M$ and $r^{*}_{\rm obs}=5200M$ for $p=7.9456$ and $p=46.062$, respectively.}
\begin{ruledtabular}
\begin{tabular}{rlcccccc}
$l$ & $m$ & $\dot{E}^{\infty}$ (FD) & $\dot{E}^{\infty}$ (TD) & rel. diff. & $\dot{L}^{\infty}$ (FD) & $\dot{L}^{\infty}$ (TD) & rel. diff.\\
\hline \hline 
&$p=7.9456$&&&&& \\ 
\hline
 2  &  1 & 8.1633e-07 & 8.1623e-07 & $<0.1\%$ & 1.8283e-05 & 1.8270e-05 & $0.1\%$   \\
    &  2 & 1.7063e-04 & 1.7051e-04 & $<0.1\%$ & 3.8215e-03 & 3.8164e-03 & $0.1\%$   \\
 3  &  1 & 2.1731e-09 & 2.1741e-09 & $<0.1\%$ & 4.8670e-08 & 4.8684e-08 & $<0.1\%$  \\
    &  2 & 2.5199e-07 & 2.5164e-07 & $0.1\%$  & 5.6439e-06 & 5.6262e-06 & $0.3\%$   \\
    &  3 & 2.5471e-05 & 2.5432e-05 & $0.1\%$  & 5.7048e-04 & 5.6878e-04 & $0.3\%$   \\
 4  &  1 & 8.3956e-13 & 8.3507e-13 & $0.2\%$  & 1.8803e-11 & 1.8692e-11 & $0.6\%$   \\
    &  2 & 2.5091e-09 & 2.4986e-09 & $0.4\%$  & 5.6195e-08 & 5.5926e-08 & $0.5\%$   \\
    &  3 & 5.7751e-08 & 5.7464e-08 & $0.5\%$  & 1.2934e-06 & 1.2933e-06 & $<0.1\%$  \\
    &  4 & 4.7256e-06 & 4.7080e-06 & $0.4\%$  & 1.0584e-04 & 1.0518e-04 & $0.6\%$   \\
 5  &  1 & 1.2594e-15 & 1.2544e-15 & $0.4\%$  & 2.8206e-14 & 2.8090e-14 & $0.4\%$   \\
    &  2 & 2.7896e-12 & 2.7587e-12 & $1.1\%$  & 6.2479e-11 & 6.1679e-11 & $1.3\%$   \\
    &  3 & 1.0933e-09 & 1.0830e-09 & $1.0\%$  & 2.4486e-08 & 2.4227e-08 & $1.1\%$   \\
    &  4 & 1.2324e-08 & 1.2193e-08 & $1.1\%$  & 2.7603e-07 & 2.7114e-07 & $1.8\%$   \\
    &  5 & 9.4563e-07 & 9.3835e-07 & $0.8\%$  & 2.1179e-05 & 2.0933e-05 & $1.2\%$   \\
\hline
& Total & 2.0317e-04 & 2.0273e-04 & $0.2\%$ & 4.5446e-03 & 4.5399e-03 & $0.1\%$  \\
\hline \hline
&$p=46.062$&&&&& \\ 
\hline
 2  &  1 & 1.8490e-11 & 1.8713e-11 & $1.2\%$ & 5.7804e-09 & 5.8497e-09 & $1.2\%$   \\
    &  2 & 2.8650e-08 & 2.8728e-08 & $0.3\%$ & 8.9566e-06 & 8.9809e-06 & $0.3\%$   \\
 3  &  1 & 7.5485e-14 & 7.7275e-14 & $2.4\%$ & 2.3598e-11 & 2.4158e-11 & $2.4\%$  \\
    &  2 & 1.0926e-12 & 1.0990e-12 & $0.6\%$ & 3.4157e-10 & 3.4359e-10 & $0.6\%$   \\
    &  3 & 8.0640e-10 & 8.0835e-10 & $0.2\%$ & 2.5210e-07 & 2.5270e-07 & $0.2\%$   \\
 4  &  1 & 9.9792e-19 & 1.0390e-18 & $4.1\%$ & 3.1191e-16 & 3.2480e-16 & $4.1\%$   \\
    &  2 & 1.6018e-14 & 1.6171e-14 & $1.0\%$ & 5.0075e-12 & 5.0555e-12 & $1.0\%$   \\
    &  3 & 4.6603e-14 & 4.6799e-14 & $0.4\%$ & 1.4569e-11 & 1.4631e-11 & $0.4\%$  \\
    &  4 & 2.7937e-11 & 2.7997e-11 & $0.2\%$ & 8.7339e-09 & 8.7525e-09 & $0.2\%$   \\
\hline
& Total & 2.9505e-04 & 2.9584e-08 & $0.3\%$ & 9.2239e-06 & 9.2486e-06 & $0.3\%$  \\
\end{tabular}
\end{ruledtabular}
\end{table}
In Table~\ref{tab:cenrg} we perform a mode by mode comparison between 
the two methods for
$p=7.9456$ and $p=46.062$.
For $p=7.9456$, the fluxes from each multipole moment calculated with the 
time-domain code agree to $1\%$ or better with the 
fluxes calculated in the frequency domain.
A similar agreement is found for $p=46.062$, with the exception of the 
$l=3$ and $m=1$, 
and $l=4$ and $m=1$ modes, for which the relative difference is $2.4\%$ and $4.1\%$, respectively.  
This results from the huge difference 
in amplitude between these modes and the dominant mode.  The stepsize 
used throughout this work is sufficient to obtain an overall relative accuracy of $1\%$, but 
it is not small enough 
to determine individual, small-amplitude modes to better than $2\thicksim 5 \%$. 
Although these modes could be resolved properly by using a smaller stepsize, 
it is not necessary for our goal of $1\%$ overall accuracy; the contributions from 
these modes to the total fluxes are six and ten orders of magnitude smaller than the 
leading-order contributions, respectively.  As such, they do not affect 
the overall accuracy of the computation; for $p=7.9456$ and $p=46.062$ 
the total fluxes calculated with the time-domain method 
agree with the frequency domain results to within $0.2\%$ and $0.3\%$, respectively.     

\begin{figure}[!htb]
\vspace*{2.5in}
\special{hscale=31 vscale=31 hoffset=-13.0 voffset=190.0
         angle=-90.0 psfile=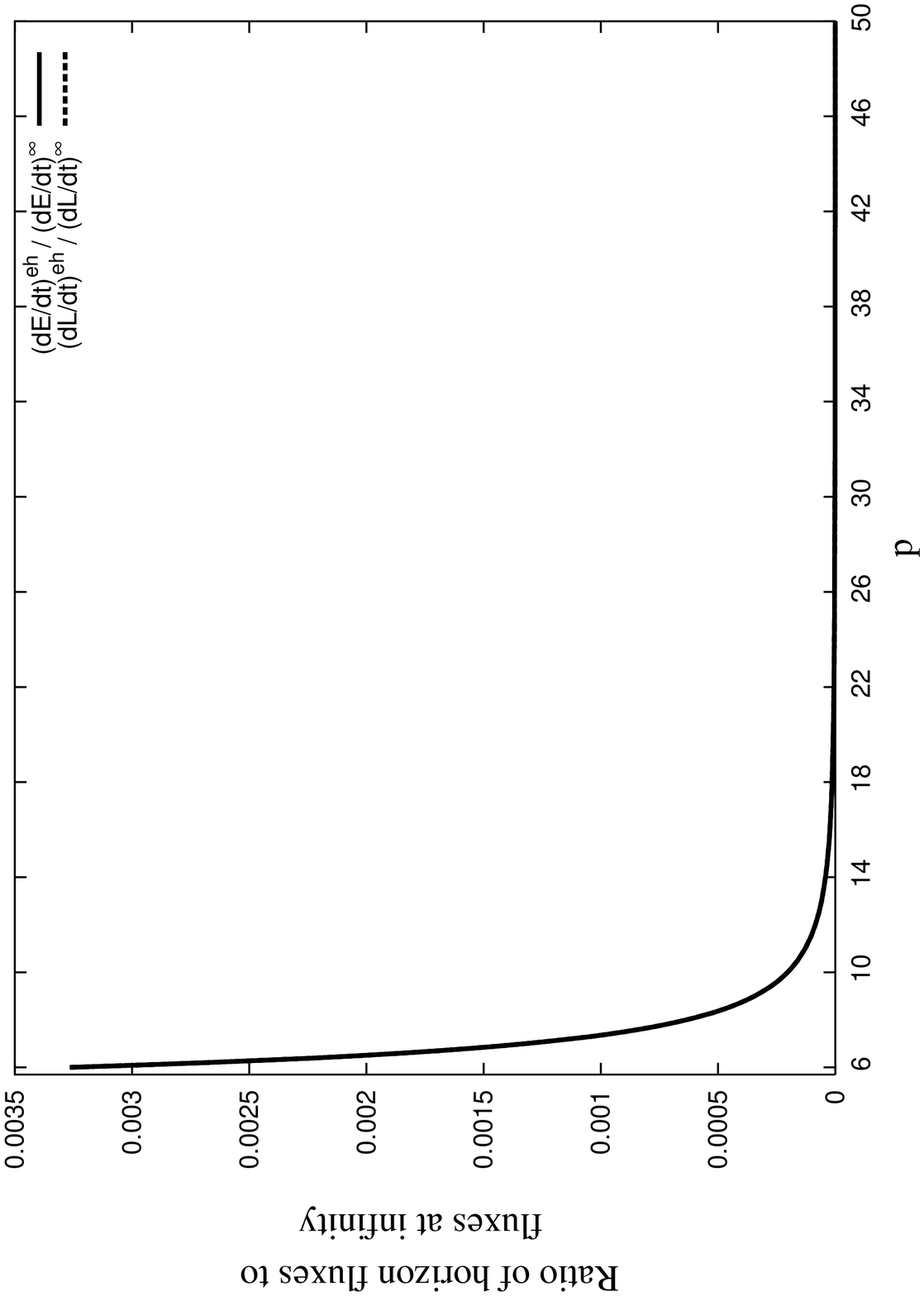}
\special{hscale=31 vscale=31 hoffset=225.0 voffset=190.0
         angle=-90.0 psfile=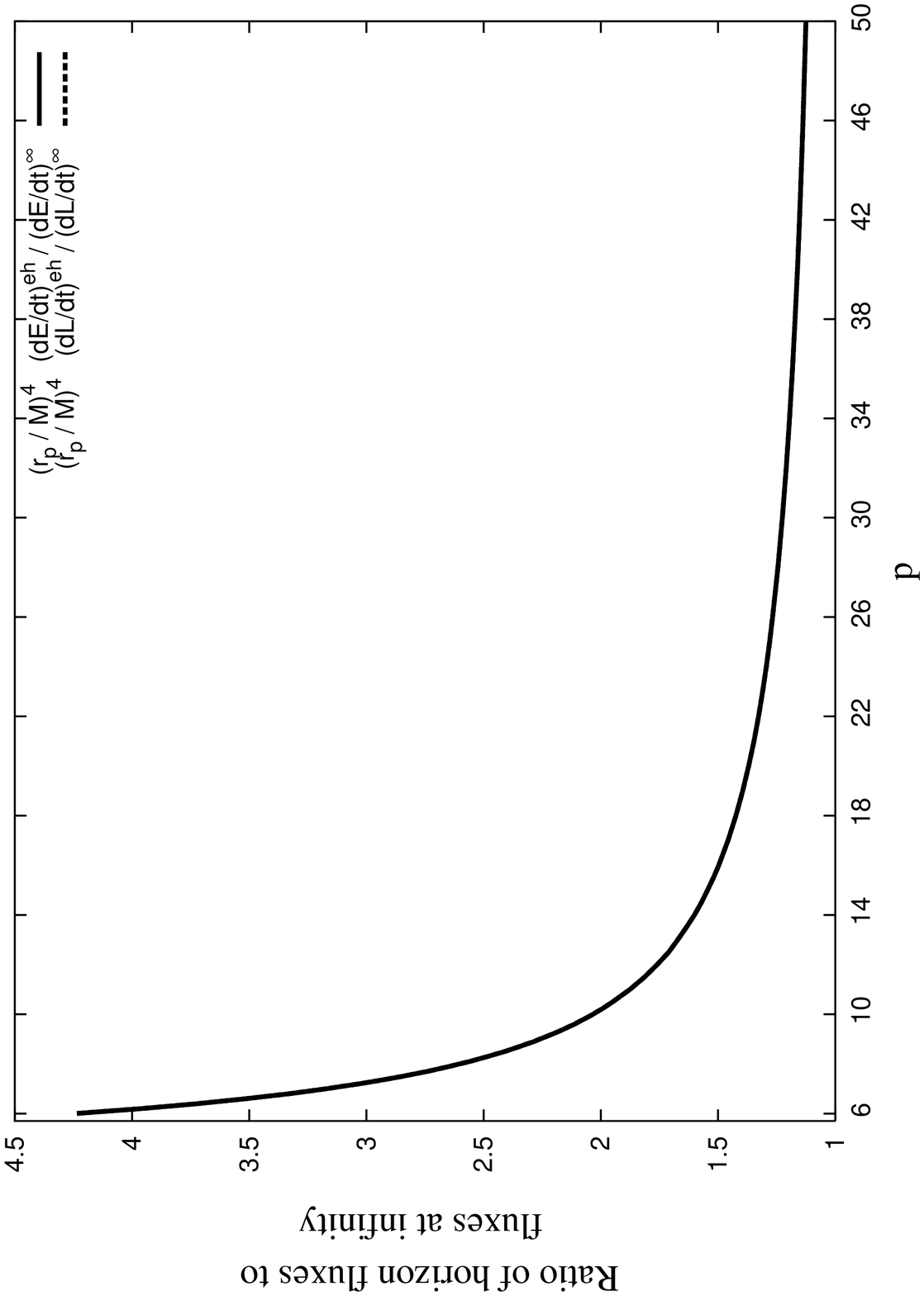}           
\caption{We display the energy and angular momentum 
fluxes through the event horizon normalized by the fluxes in the radiation zone.
Even for highly relativistic motion, the horizon fluxes contribute less than $0.4\%$ 
of the total fluxes.  For circular orbits, the theoretical prediction is that 
$\dot{E}^{\rm eh}/\dot{E}^{\infty}=\dot{L}^{\rm eh}/\dot{L}^{\infty}$.  
Numerically, this relation is only approximate, 
but nevertheless the two curves are indistinguishable. 
The right panel displays these ratios 
normalized by $(r_{p}/M)^{-4}$, the weak-field and slow-motion approximation.} \label{fig:cehflux}
\end{figure}
Black hole absorption was calculated 
in a weak-field and slow-motion approximation for a particle in circular orbit 
by E.~Poisson and M.~Sasaki~\cite{Sasaki} and K.~Alvi~\cite{Alvi} who 
showed that it gives rise to a $v^{8}$ correction to the quadrupole formula:
$\dot{E}^{{\rm eh}}/\dot{E}_{Q}=v^{8}=\dot{L}^{{\rm eh}}/\dot{L}_{Q}$, 
where $v=p^{-1/2}=(M/r_{p})^{1/2}$ is the orbital velocity.  
The time-domain method allows black hole 
absorption to be calculated for arbitrary geodesics. 
In particular, for circular orbits 
our results show that even when $v\sim 0.4$ and the particle travels in a region of strong gravitational field, 
the amount of energy and angular momentum absorbed by the black hole 
is always a small correction to the total fluxes.
For highly relativistic motion, this 
never grows large enough to contribute 
more than $0.4\%$ of the total fluxes (see left panel of Fig.~\ref{fig:cehflux}). 
For the purpose of calculating total fluxes with an overall accuracy of $1\%$, 
black hole absorption can safely be ignored.  
The right panel displays the ratio of horizon fluxes to the fluxes at infinity, 
normalized by $(M/r_{p})^{4}$, the weak-field and slow-motion approximation.  
As expected for circular motion, the normalized ratio for energy and angular 
momentum are equal to each other, and they approach 1 for large $p$.

We estimate the accuracy with which black hole absorption 
can be determined using the time-domain method to be $5\%$.  This estimate is based 
on the following argument.  For small amplitude modes, the accuracy 
with which their 
contribution to the total fluxes can be determined is limited by errors originating 
from the discretization of Eq.~(\ref{eqn:wave}) and the 
finite stepsize used in the 
numerical evolution.  Based on the accuracy of 
the $l=4$ and $m=1$ mode for $p=46.062$ in Table~\ref{tab:cenrg}, the error is seen to be $\lesssim 5\%$ 
for modes whose contribution is ten orders of magnitude smaller than the dominant mode.  
For the range of $p$ values considered in this paper, 
black hole absorption is at most seven orders of magnitude smaller 
than the dominant contribution. (This is evaluated using the $(M/r_{p})^{4}$ 
relation at $r_{p}=50 M$, the value at which black hole absorption is least significant.)
It is then safe to assume that fluxes through the event horizon are determined with an accuracy $<5\%$.  
(For values of $p$ close to $6+2e$, black hole absorption is more important 
and therefore more accurately determined.)

%%%%%%%%%%%%%%%%%%%%%%%%%%%%%%%%%%%%%%%%%%%%%%%%%%%%%%%%%%%%%%%%%%%%%%%%%%%%
%%%%%%%%%%%%%%%%%%%%%%%%%%%%%%%%%%%%%%%%%%%%%%%%%%%%%%%%%%%%%%%%%%%%%%%%%%%%

\subsection{Eccentric orbits} \label{sec:eccentric}
\begin{figure}[!htb]
\vspace*{2.5in}
\special{hscale=31 vscale=31 hoffset=-13.0 voffset=190.0
         angle=-90.0 psfile=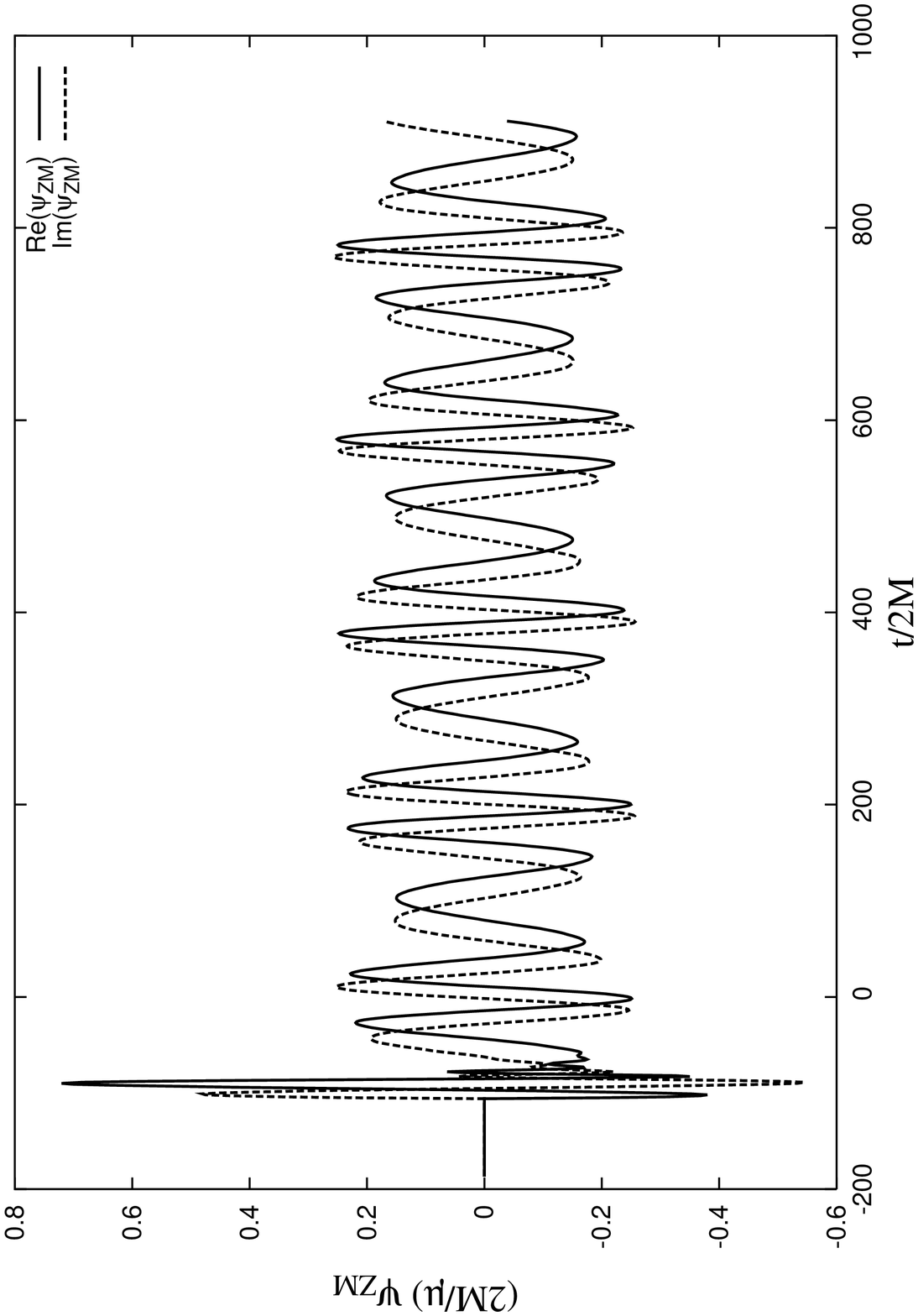}
\special{hscale=31 vscale=31 hoffset=225.0 voffset=190.0
         angle=-90.0 psfile=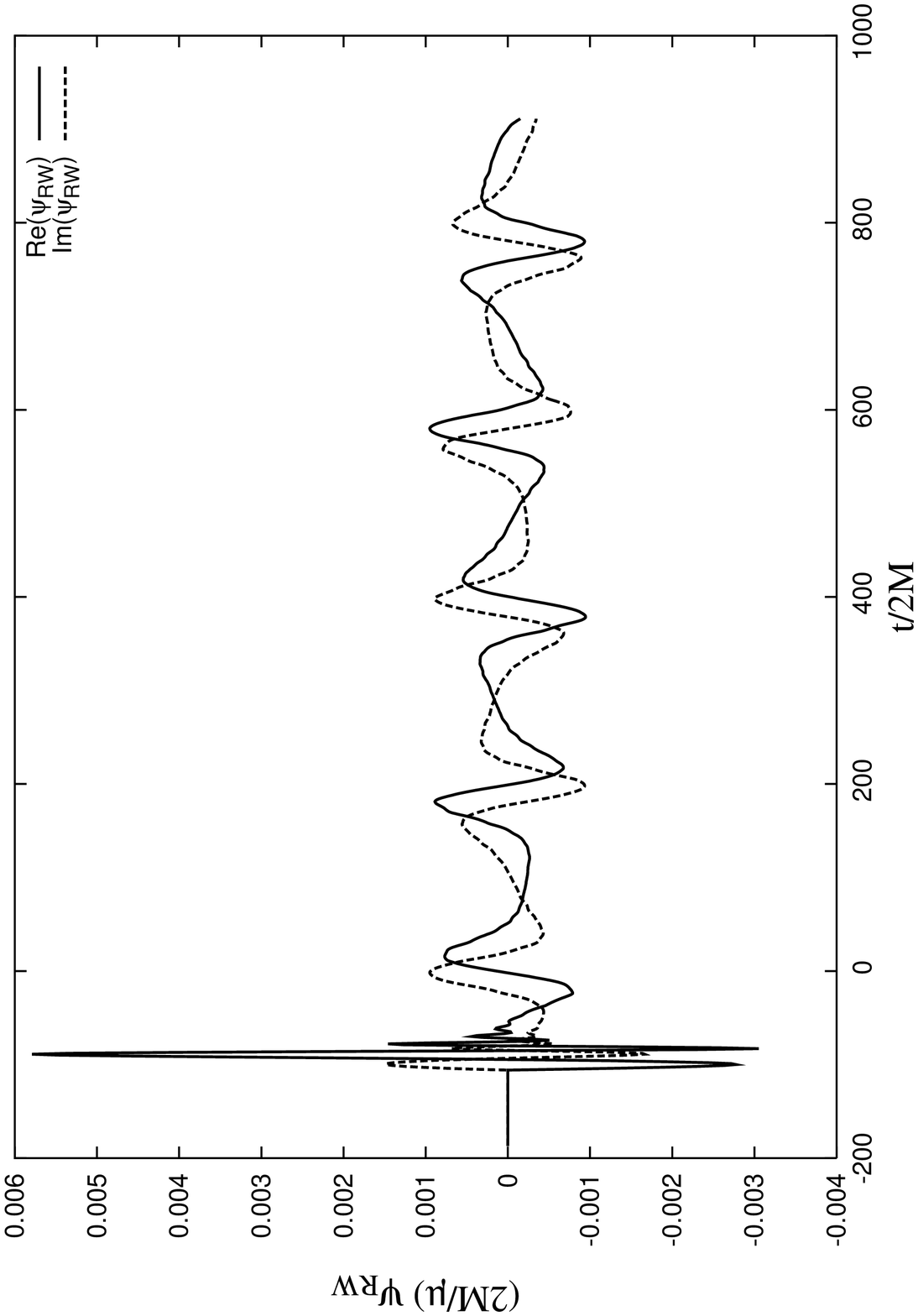}
\caption{The Zerilli-Moncrief (left, $l=2, m=2$) 
and Regge-Wheeler (right, $l=2, m=1$) 
functions for $p=12$ and $e=0.2$.  As in the case 
of circular orbits, early times are dominated by 
the initial data content.}\label{fig:normalpe}
\end{figure}

\begin{figure}[!htb]
\vspace*{2.5in}
\special{hscale=31 vscale=31 hoffset=-13.0 voffset=190.0
         angle=-90.0 psfile=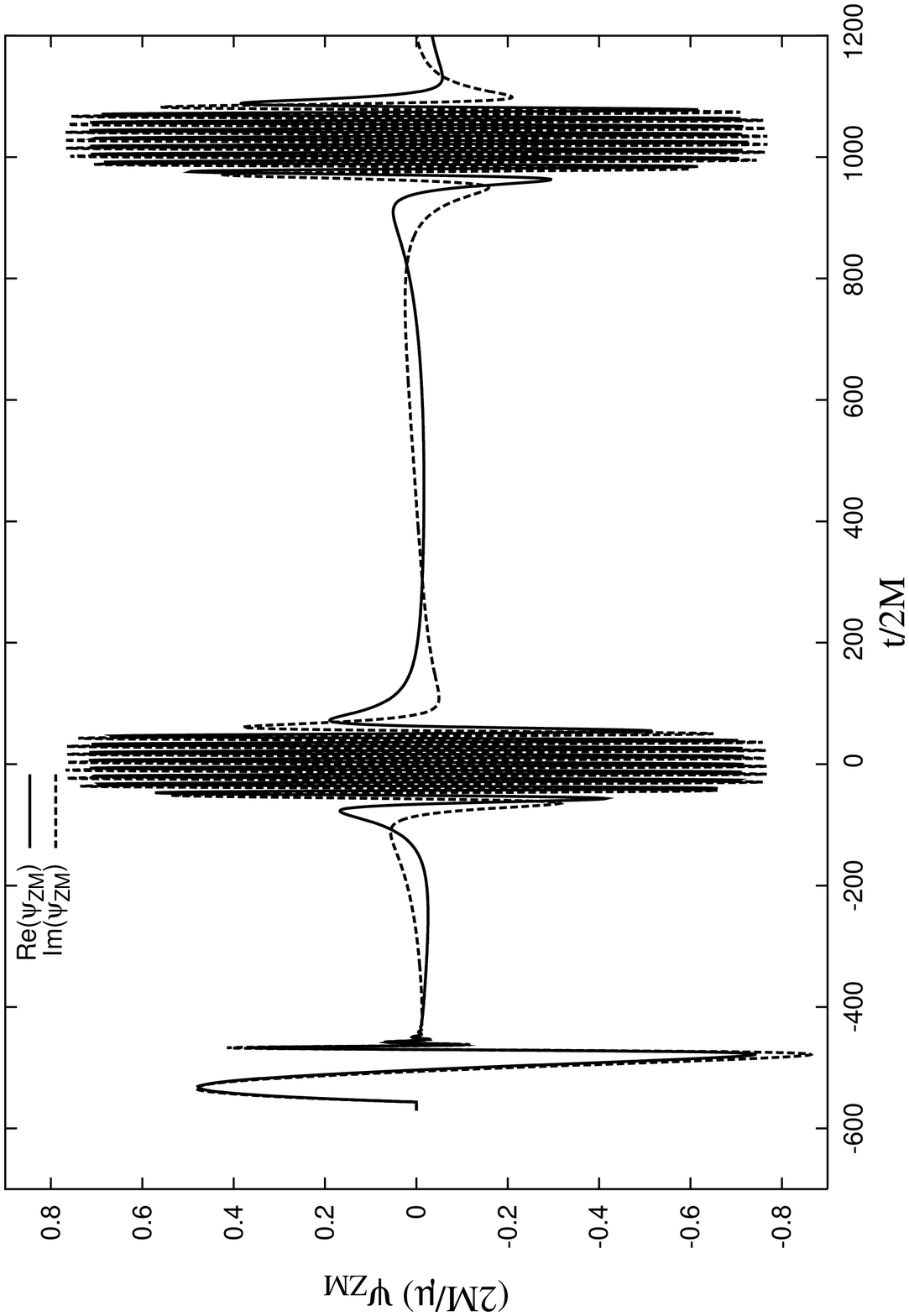}
\special{hscale=31 vscale=31 hoffset=225.0 voffset=190.0
         angle=-90.0 psfile=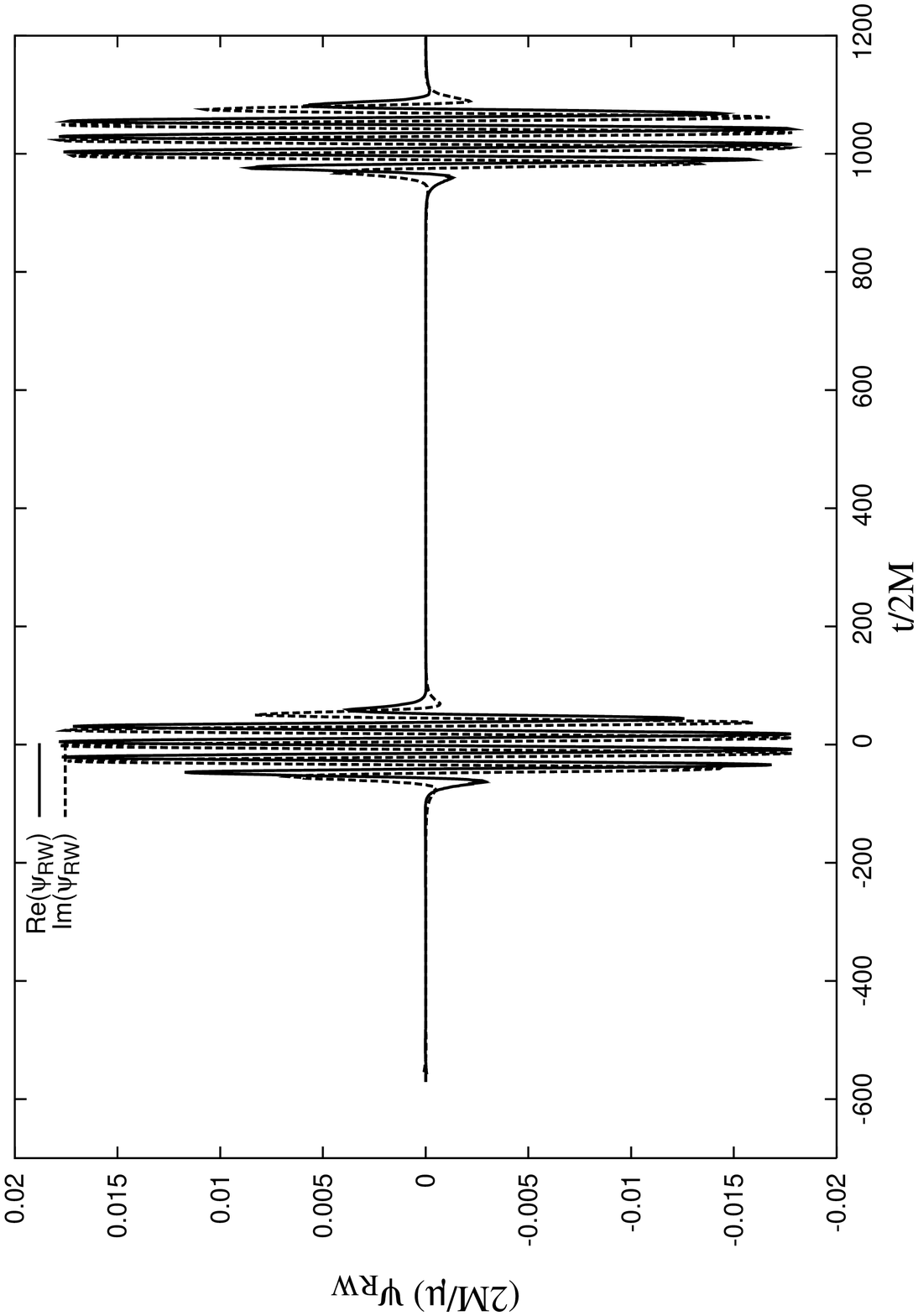}
\caption{The Zerilli-Moncrief (left, $l=2, m=2$) and 
Regge-Wheeler (right, $l=2, m=1$) 
functions for $p=7.801$ and $e=0.9$.  As in 
the case of circular orbits, early times are dominated by 
the initial data content. The radiation occurs in short bursts  
when the particle approaches the periastron.  This is typical 
of the zoom-whirl behaviour studied in~\cite{Kennefick}.}\label{fig:smallpe}
\end{figure}

For eccentric orbits, $0<e<1$, and the radial motion is bounded 
by the periastron $r_{p}|_{\rm min}=pM/(1+e)$
and the apastron $r_{p}|_{max}=pM/(1-e)$.  
In Fig.~\ref{fig:normalpe} and Fig~\ref{fig:smallpe}  
we display waveforms for two cases: $p=12$ and $e=0.2$, as 
well as $p=7.801$ and $e=0.9$. 

This type of orbital motion generates 
gravitational waveforms that are different in nature and in frequency content from
circular orbits.  Rather than being emitted uniformly along the orbit, the radiation is now emitted 
preferably at periastron. As the eccentricity increases the radiation is emitted in short bursts occurring 
near periastron.  In these situations a time-domain approach is far more efficient than a frequency-domain approach. 
The reason is that in order to correctly calculate the waveforms in the frequency domain, a large number of individual frequencies 
(harmonics of the radial and azimuthal frequencies) are required, and summing 
over them can be hugely expensive. By contrast, a time-domain method handles all frequencies simultaneously.

The fluxes are calculated over a number of wave cycles according to
\begin{eqnarray} \label{eqn:averaged}
<\dot{E}> &=& \frac{1}{T}\int_{0}^{T} \dot{E} \ dt, 
\end{eqnarray}
where $T$ is a few ($>3$) radial periods; a similar expression holds for $<\dot{L}>$. 
To obtain a quantitative idea of the relative 
accuracy of the time-domain method for eccentric orbits, 
we compute the fluxes for 
two points in the $p$-$e$ plane and compare our calculations 
with Cutler {\it et al.}~\cite{Cutler}: i) $p=7.50478$ 
and $e=0.188917$, 
and ii) $p=8.75455$ and $e=0.764124$.  The results are displayed 
in Table~\ref{tab:EL_ecc}.  For small eccentricities, e.g. case (i), the agreement is similar 
to the agreement achieved for circular orbits. For large eccentricities, e.g. case (ii), 
the agreement is $\thicksim 2\%$. Because eccentric orbits in Schwarzschild are characterized by two 
incommensurate frequencies, the gravitational waveforms are quasi-periodic.  
By working in the frequency domain, Cutler {\it et al.} were able to formally average  
their fluxes over an infinite time.  It is not, of course, possible to perform such an 
average in the time domain.  
Rather, for high eccentricities, the fluxes are averaged over a limited number of radial cycles ($\sim 3$). 
This difference in averaging the fluxes is the
most likely source of disagreement 
between time-domain and frequency-domain calculations 
for case (ii).  For case (i), this is not as much of an issue, 
since the radial period is short enough to allow the time average to be performed 
over 10 cycles or more.
\begin{table}
\caption{Comparison of averaged fluxes for eccentric orbits with Cutler 
{\it et al.} for two points in the $p$-$e$ plane~\cite{Cutler}.  
The two cases presented are: i) $p=7.50478$ and $e=0.188917$, 
and ii) $p=8.75455$ and $e=0.764124$.\label{tab:EL_ecc}}
\begin{ruledtabular}
\begin{tabular}{rccccc}
&case && Cutler {\it et al.}& time domain & rel. diff.\\
\hline \hline
i) & $p=7.50478$ & $<\dot{E}>$ & 3.1680e-04 & 3.1770e-04 & $0.3\%$  \\
   & $e=0.188917$& $<\dot{L}>$ & 5.9656e-03 & 5.9329e-03 & $0.5\%$  \\
\hline
ii) & $p=8.75455$ & $<\dot{E}>$ & 2.1008e-04 & 2.1484e-04 & $2.3\%$ \\
    & $e=0.764124$& $<\dot{L}>$ & 2.7503e-03 & 2.7932e-03 & $1.6\%$  \\
\end{tabular} 
\end{ruledtabular}
\end{table}

\begin{figure}[!htb]
\vspace*{2.5in}
\special{hscale=31 vscale=31 hoffset=90.0 voffset=190.0
         angle=-90.0 psfile=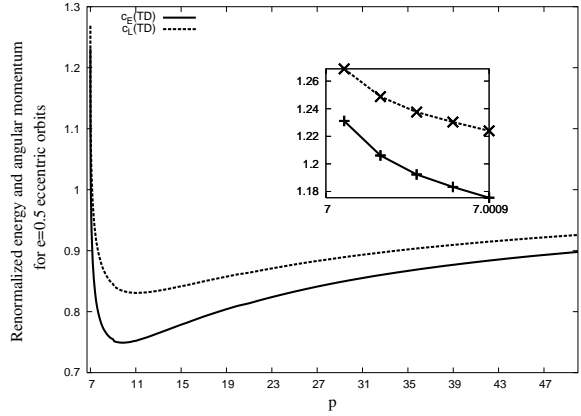}
\caption{Coefficients $c_{E}$ and $c_{L}$ for 
the energy and angular momentum radiated as functions of $p$ for $e=0.5$ eccentric orbits.  
Near the last stable orbit ($p=7$), 
$c_{E}$ approaches 1.24, while $c_{L}$ approaches 1.26.}\label{fig:eccFlux}
\end{figure}
Finally, we calculate the total energy and angular momentum emitted 
during one radial period as functions of $p$ for $e=0.5$.  
We express the total energy and angular momentum radiated to infinity, as calculated 
from the time domain code, as
\begin{eqnarray} \label{eqn:EL_ecc}
E_{GR}(p,e) &=& c_{E} \left[E_{Q}(p,e)+(N-1)E_{Q}(p/(1+e),0)\right], \nonumber \\
L_{GR}(p,e) &=& c_{L} \left[L_{Q}(p,e)+(N-1)L_{Q}(p/(1+e),0)\right],
\end{eqnarray}
where we use $E_{GR}=P(p,e)<\dot{E}>$, $P(p,e)$ is the radial period of 
the orbit obtained by integrating Eq.~(\ref{eqn:dt}) over $0\leq \chi \leq 2\pi$,
$<\dot{E}>$ is given by Eq.~(\ref{eqn:averaged}), 
$N=N(p,e)$ is given by Eq.~(\ref{eqn:Ncycle}) with $e=0.5$, 
$\dot{E}_{Q}(p,e)$ and $\dot{L}_{Q}(p,e)$ are given by Eq.~(\ref{eqn:QuadApp}),  
and $c_{E}$ and $c_{L}$ are parameters 
that stay close to 1 for all values of $p$.
In Fig.~\ref{fig:eccFlux} we display $c_{E}$ and $c_{L}$ 
as functions of $p$ for $e=0.5$. The coefficient $c_{E}$ is close to 0.9 for large $p$ and approaches 
1.24 for $p$ near 7.  Similarly, the coefficient $c_{L}$ stays close to 0.95 for large $p$ and 
approaches 1.26 for $p$ near 7.   
The formulae above for the total energy and angular momentum radiated by a particle in eccentric orbit 
are justified by the fact that they have the correct limiting behaviour both 
for $p$ large and for $p\rightarrow 6+2e$.
For large $p$, the total energy and angular momentum radiated by a particle on an eccentric 
orbit are well approximated by the quadrupole formulae given by Eq.~(\ref{eqn:QuadAppTot}).  
In this limit, $N\rightarrow 1$ and Eq.~(\ref{eqn:EL_ecc}) produces the 
correct approximate energy and angular momentum radiated.  When $p\rightarrow 6+2e$, 
the particle orbits the black hole for a number $N-1$ of quasi-circular 
orbits whose radius is equal to the periastron radius $r_{p}=Mp/(1+e)$.  
In this limit $N$ is large and the second term of Eq.~(\ref{eqn:EL_ecc}) dominates the energy and angular momentum radiated.  
This term corresponds to the energy and angular momentum radiated after $N$ such quasi-circular orbits

\begin{figure}[!htb]
\vspace*{2.5in}
\special{hscale=31 vscale=31 hoffset=-13.0 voffset=190.0
         angle=-90.0 psfile=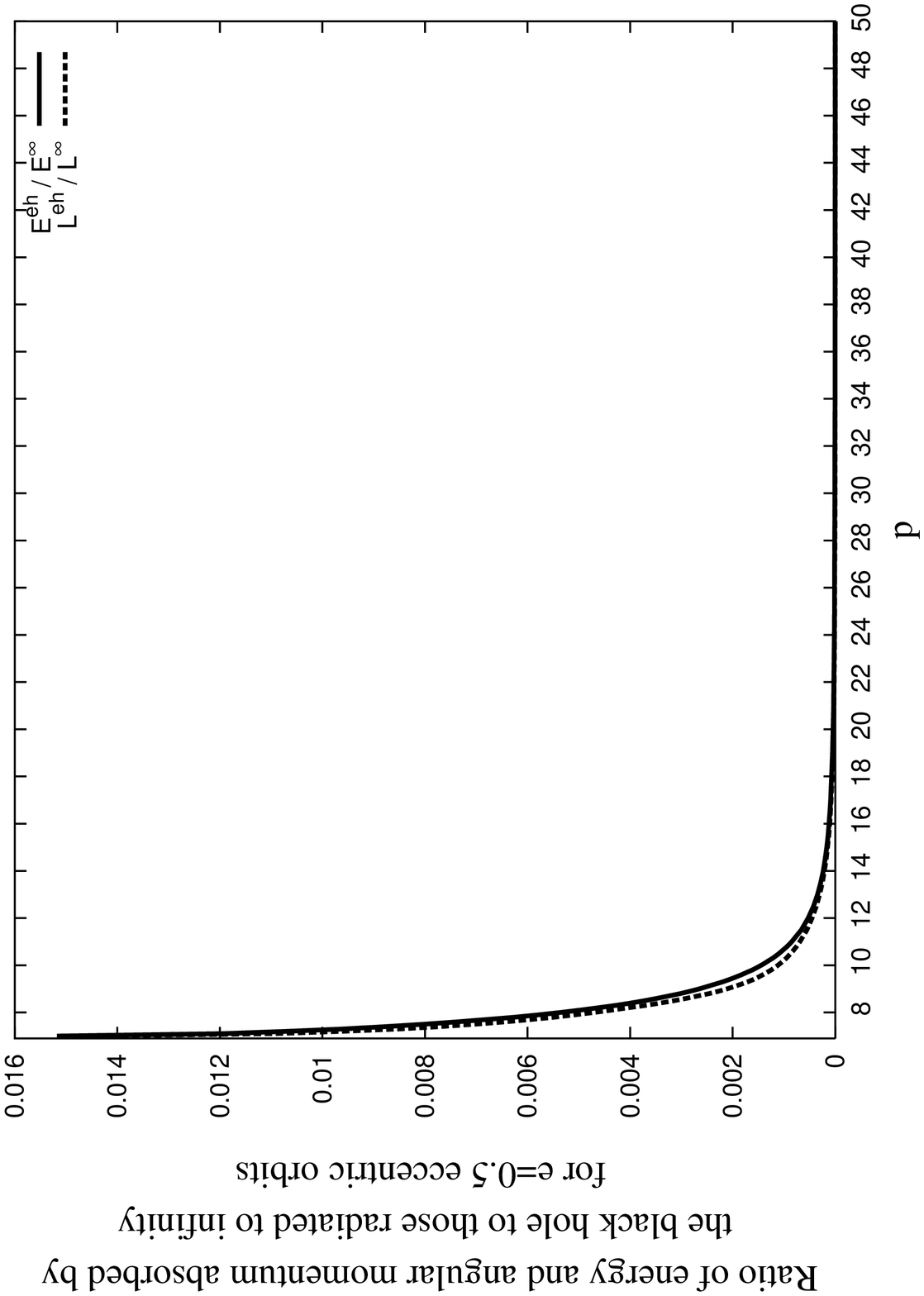}
\special{hscale=31 vscale=31 hoffset=225.0 voffset=190.0
         angle=-90.0 psfile=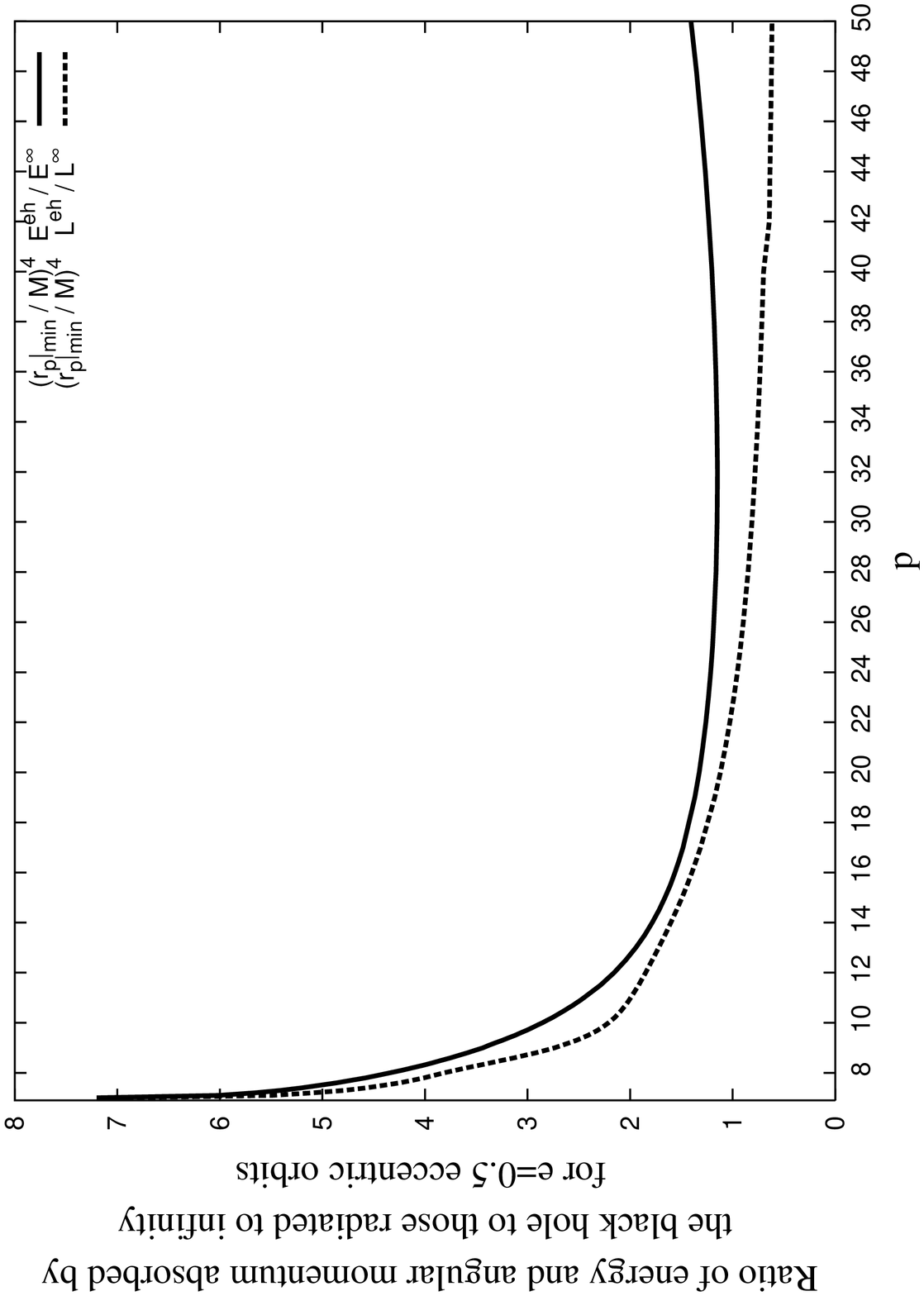}
\caption{Black hole absorption for a particle in an $e=0.5$ eccentric orbit.  
The absorption of both energy and angular momentum is negligible until the particle 
reaches $p\approx 7.3$, at which point it contributes approximately $1\sim 2\%$ of 
the total fluxes; these are eccentric orbits whose periastron is smaller than $4.9M$.
The right panel displays the same ratio normalized by $(M/r_{p}|_{\rm min})^{4}$. 
}\label{fig:eccEHflux}
\end{figure}
The frequency of the radiation emitted by the orbiting particle increases as 
the periastron of the orbit becomes smaller.  Since for a given eccentricity $e$, 
the periastron is proportional to $p$,  the frequency of the radiation increases with decreasing 
$p$.  Because the potential barrier around the 
Schwarzschild black hole is less opaque to high-frequency gravitational waves, 
we expect an increase in black hole absorption with a 
decrease in $p$.  This is confirmed numerically for $e=0.5$ 
and displayed in Fig.~\ref{fig:eccEHflux}.
For $p \lesssim 7.3$ ($r_{p}\approx 4.9 M$), the absorption of energy and 
angular momentum by the black hole contributes  more than $1\%$ of 
the total fluxes, while for $p\gtrsim 7.3$ it 
contributes less than $1\%$ 
and can be ignored when determining 
the total fluxes.  
In the right panel of the figure, we display black hole absorption for $e=0.5$,  
normalized by $(M/r_{p}|_{\rm min})^{4}$, where $r_{p}|_{\rm min}=Mp/(1+e)$ 
is the periastron distance.  This is the correction 
expected from black hole absorption for a particle in {\em circular} 
orbit at $r_{p}|_{\rm min}$.  We use this normalization here because black hole absorption for generic orbits 
has not been calculated analytically.
For large $p$, black hole absorption 
for $e=0.5$ does not seem to converge toward the slow-motion 
and weak-field approximation for circular orbits.  
The normalized energy stays above 1, 
while the normalized angular momentum curve stays below 1.   
But because the relation $dE=\Omega dL$ used in deriving black hole absorption 
for circular orbit does not hold in general, there is no reason to believe that 
$(M/r_{p}|_{\rm min})^{4}$ should hold for generic orbits.
Determining the differences in black hole absorption due to a finite eccentricity in a weak-field and slow-motion 
approximation would require a more detailed analysis than ours, since it is in this 
regime that our determination of black hole absorption is the least accurate.

\begin{figure}[!htb]
\vspace*{2.5in}
\special{hscale=31 vscale=31 hoffset=-13.0 voffset=190.0
         angle=-90.0 psfile=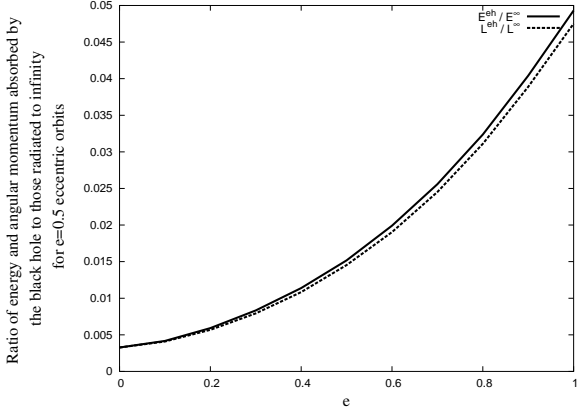}
\caption{Displayed are $E^{\rm eh}/E^{\infty}$ and $L^{\rm eh}/L^{\infty}$ 
as functions of eccentricity 
along the curve $p=6.001+2e$. Because of the decrease in periastron distance 
with increasing $e$, black hole absorption increases with $e$.  
A good approximation to these curves is given by 
$E^{\rm eh}/E^{\infty}=(1+14e^2)\left(E^{\rm eh}/E^{\infty}\right)|_{e=0}$.
}\label{fig:ecc_var}
\end{figure}
For radiation emitted by a particle 
whose orbital parameters are $p=6+2e$ and $0\leq e \leq 1$, the argument 
relating black hole absorption to the orbital separation 
suggests that black hole absorption 
should be an increasing function of $e$ along the line $p=6+2e$ ($r_{p}|_{\rm min}$ is a 
decreasing function of $e$ along this line).  
It then comes as no surprise that numerical results displayed in 
Fig.~\ref{fig:ecc_var} support this assertion (we used $p=6.001+2e$). 
Along this line, the radiation is emitted principally  at periastron, where the orbit 
is quasi-circular.  The relation $dE=\Omega dL$, where 
$\Omega =(M/r_{p}|_{\rm min})^{1/2}$ is the angular velocity of a particle 
on a circular orbit at $r_{p}|_{\rm min}$, holds approximately and we find 
$E^{\rm eh}/E^{\infty}\approx L^{\rm eh}/L^{\infty}$. 

%%%%%%%%%%%%%%%%%%%%%%%%%%%%%%%%%%%%%%%%%%%%%%%%%%%%%%%%%%%%%%%%%%%%%%%%%%%%
%%%%%%%%%%%%%%%%%%%%%%%%%%%%%%%%%%%%%%%%%%%%%%%%%%%%%%%%%%%%%%%%%%%%%%%%%%%%

\subsection{Parabolic orbits} \label{sec:parabolic}
\begin{figure}[!htb]
\vspace*{2.5in}
\special{hscale=31 vscale=31 hoffset=-13.0 voffset=190.0
         angle=-90.0 psfile=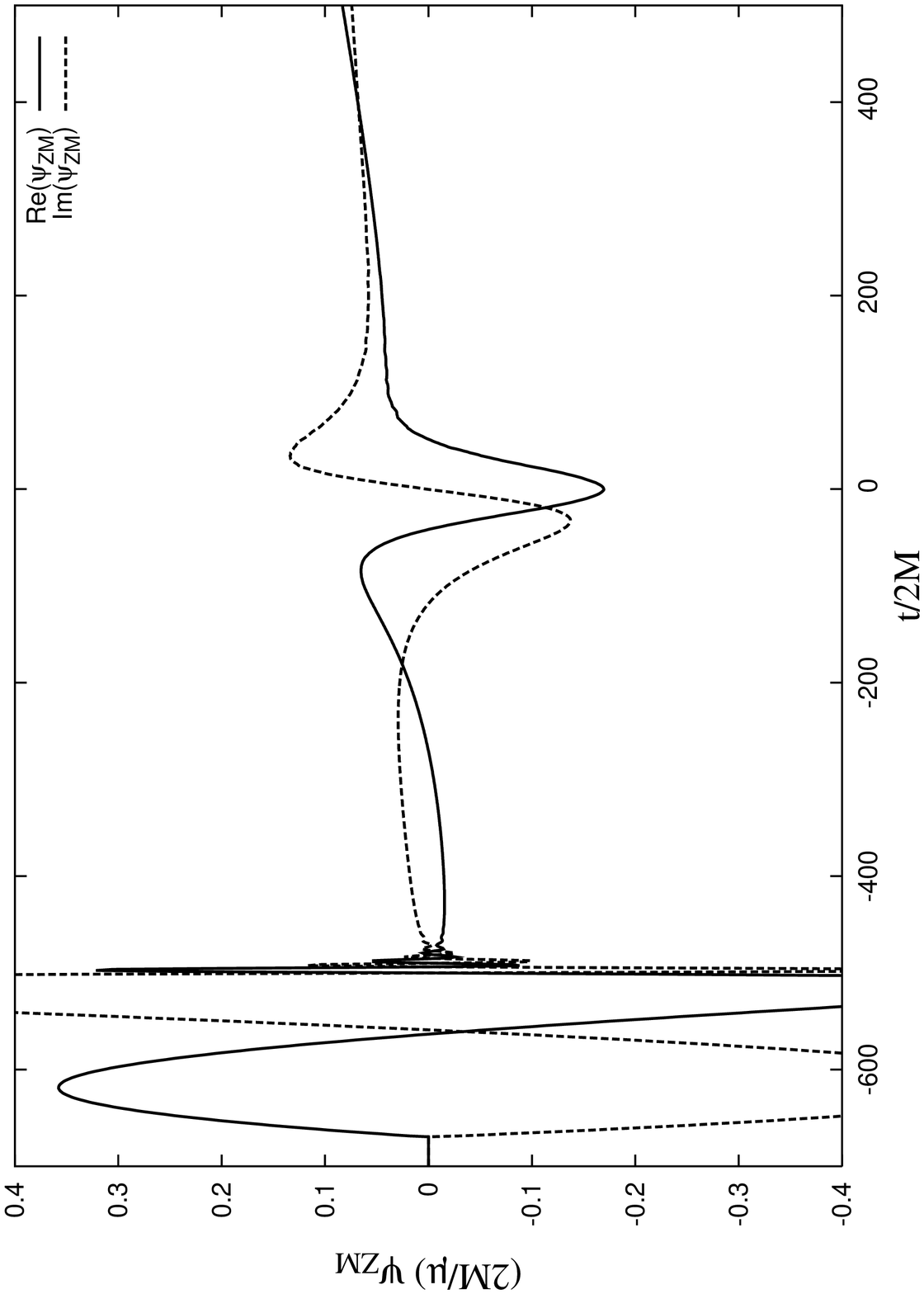}
\special{hscale=31 vscale=31 hoffset=225.0 voffset=190.0
         angle=-90.0 psfile=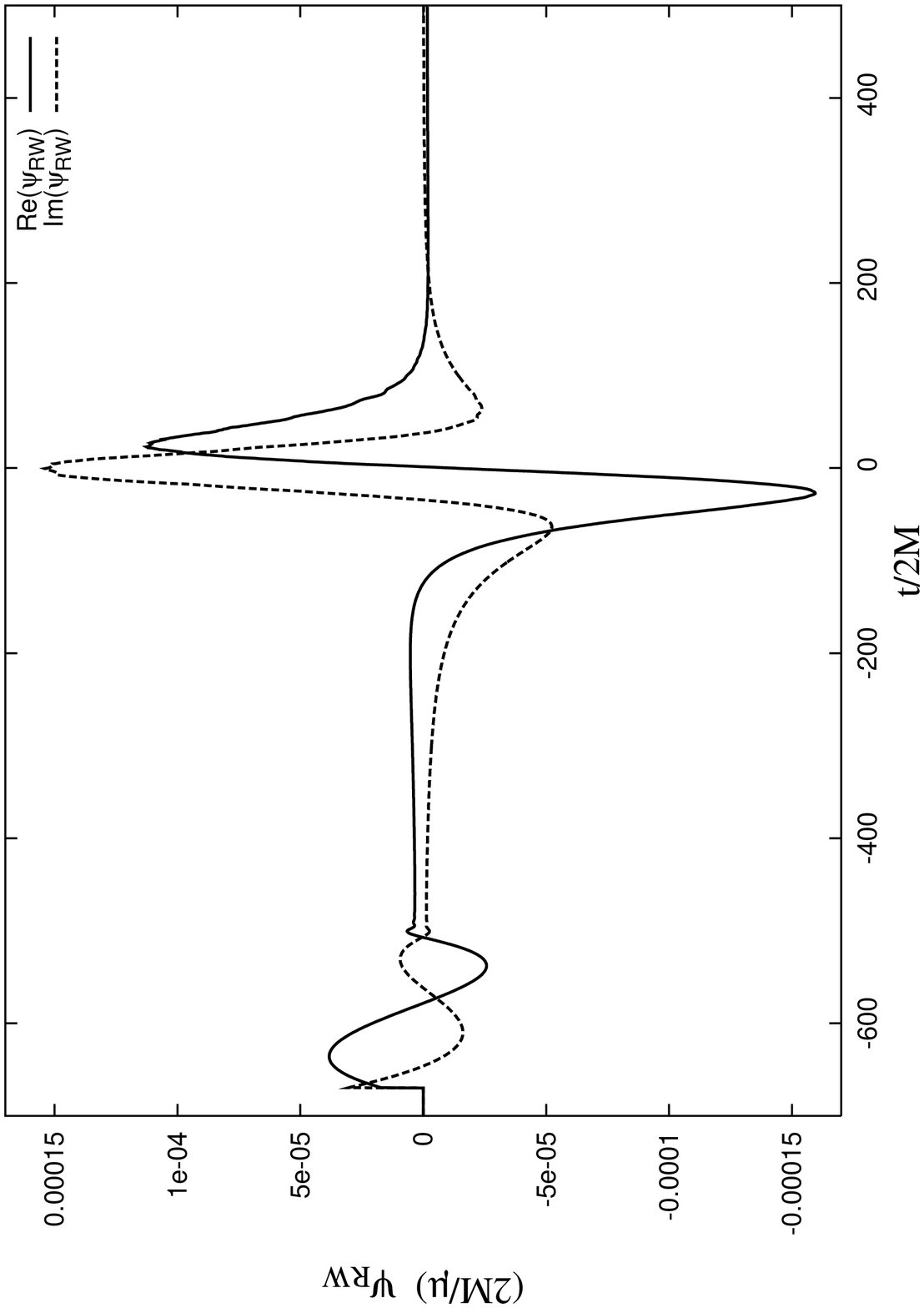}
\caption{Displayed are $\psi_{\rm ZM}$ (left, $l=2, m=2$) 
and $\psi_{\rm RW}$ (right, $l=2, m=1$) 
as functions of time for $e=1$ and $p=40$.  As in the case of 
circular orbits, early times are dominated by 
the initial data content. Total energy and angular momentum  
are calculated between $-300 \leq t/2M \leq 300$.}\label{fig:pfar}
\end{figure}
\begin{figure}[!htb]
\vspace*{2.5in}
\special{hscale=31 vscale=31 hoffset=-13.0 voffset=190.0
         angle=-90.0 psfile=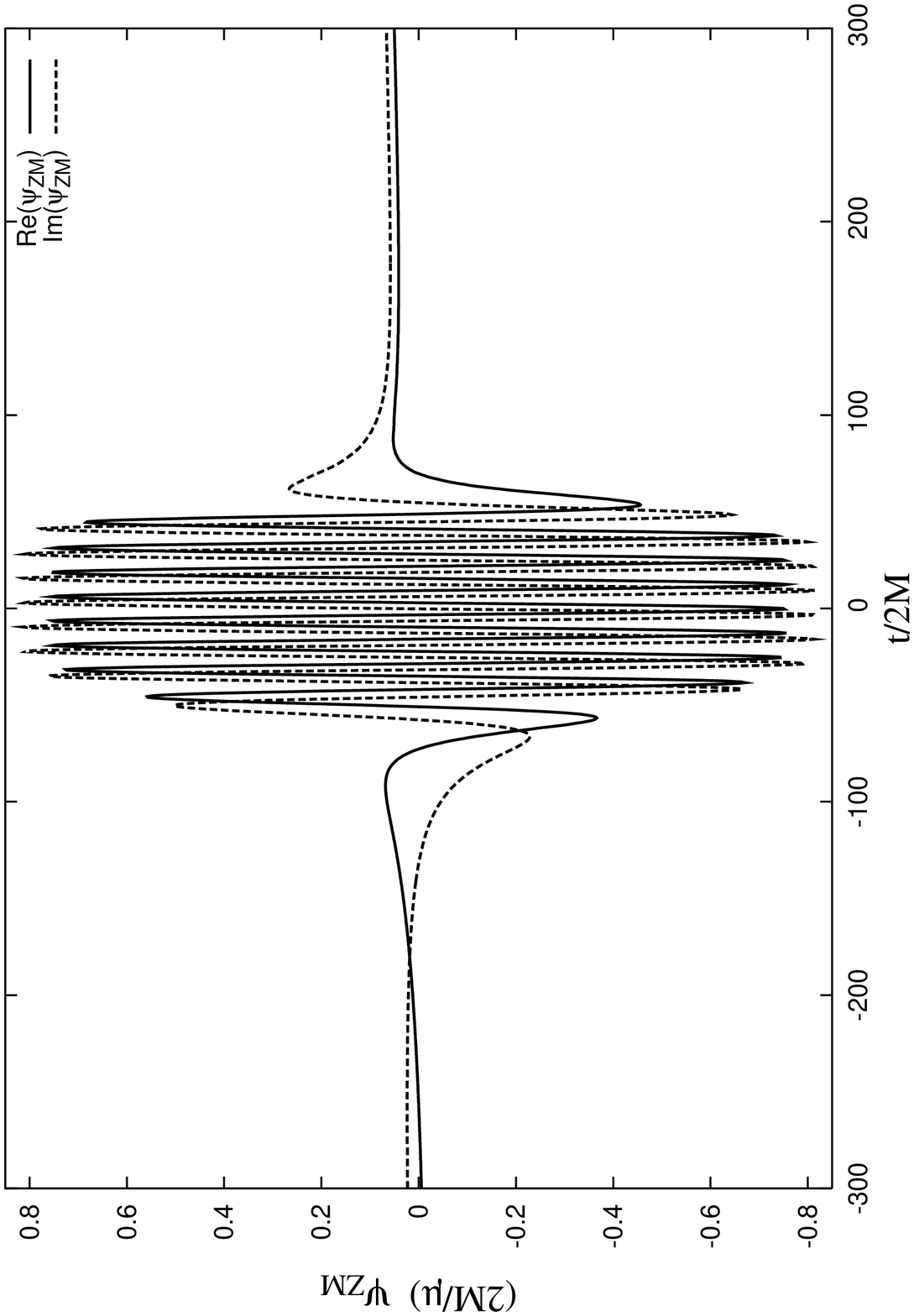}
\special{hscale=31 vscale=31 hoffset=225.0 voffset=190.0
         angle=-90.0 psfile=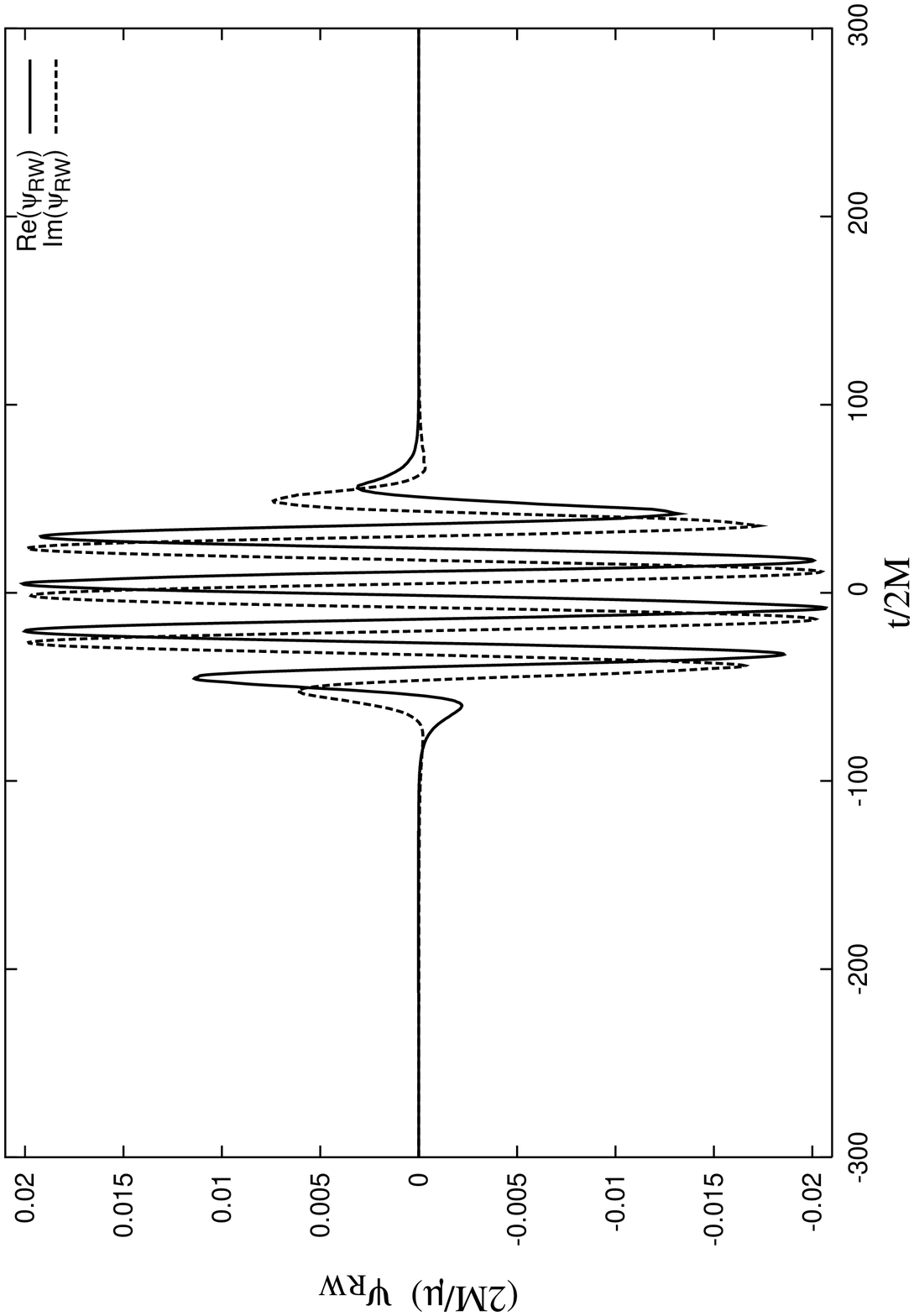}
\caption{Displayed are $\psi_{\rm ZM}$ (left, $l=2, m=2$) 
and $\psi_{\rm RW}$ (right, $l=2, m=1$) 
as functions of time for $e=1$ and $p=8.001$. Early times, 
where the choice of initial data dominates, 
are not displayed in order to make the $t=0$ region 
clearly visible. The energy and angular momentum fluxes are integrated 
between $-300 \leq t/2M \leq 300$ to obtain the total energy 
and angular momentum radiated.}\label{fig:pclose}
\end{figure}
Particles on a parabolic trajectory have $e=1$ (equivalently $\tilde{E}=1$), 
and $p$ specifies the value of 
the periastron: $r_p|_{\rm min}= M p/2$ with $p>8$. 
For large values of $p$, the particle does not spend much 
time around $r_{p}|_{\rm min}$, the position where the 
radiation is maximum; the waveforms have a simple structure 
around $t=0$, time at which the radiation emitted at $r_{p}|_{\rm min}$ 
reaches an observer at $r^{*}_{\rm obs}$. 
This is displayed for even and odd modes in Fig.~\ref{fig:pfar}.  
In contrast, when $p$ approaches its minimum value 
($p_{\rm min} \gtrsim 8$), the particle circles the black hole 
for a number $N$ of cycles.  Because $N$ diverges at $p=8$, we get 
the zoom-whirl behaviour displayed in Fig.~\ref{fig:orbits}~\cite{Kennefick}.  
The quasi-circular nature of the motion when $r_{p}$ 
approaches $r_{p}|_{\rm min}$ results in a number of 
oscillations in the waveforms; these occur near $t=0$ 
for the observer at $r^{*}_{\rm obs}$, and are 
displayed in Fig.~\ref{fig:pclose} for $p=8.001$.

\begin{figure}[!htb]
\vspace*{2.5in}
\special{hscale=31 vscale=31 hoffset=90.0 voffset=190.0
         angle=-90.0 psfile=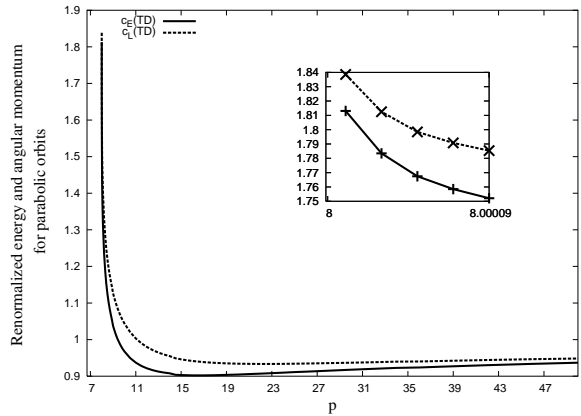}
\caption{Coefficients $c_{E}$ and $c_{L}$ for the total energy and angular momentum radiated 
as functions of $p$ for a particle in parabolic orbit.
Near $p=8$, $c_{E}$ approaches 1.81 
while $c_{L}$ approaches 1.84.}\label{fig:pde}
\end{figure}
In similarity with eccentric orbits, we express the 
numerically calculated energy and angular momentum radiated as
\begin{eqnarray}
E_{GR}(p,e) &=& c_{E} \left[E_{Q}(p,e)+(N-1)E_{Q}(p/(1+e),0)\right], \nonumber \\
L_{GR}(p,e) &=& c_{L} \left[L_{Q}(p,e)+(N-1)L_{Q}(p/(1+e),0)\right],
\end{eqnarray}
where $N=N(p,e)$ is given by Eq.~(\ref{eqn:Ncycle}) with $e=1$, 
$\dot{E}_{Q}(e,p)$ and $\dot{L}_{Q}(e,p)$ are given by Eq.~(\ref{eqn:QuadApp}), 
and $c_{E}$ and $c_{L}$ are again parameters 
that stay close to 1 for all $p$.  For parabolic orbits, the total energy and angular momentum 
are computed using Eqs.~(\ref{eqn:Etot}) and (\ref{eqn:Ltot}) as 
\begin{eqnarray}
E^{\infty}_{GR} &=& \int_{-T}^{T} \dot{E}^{\infty}\ dt, \nonumber \\
L^{\infty}_{GR} &=& \int_{-T}^{T} \dot{L}^{\infty}\ dt, 
\end{eqnarray}
for $T$ large (we used $T=300 (2M)$).  
In Fig.~\ref{fig:pde}, 
we display $c_{E}$ and $c_{L}$ for parabolic orbits.  
These quantities are close to 1 for large $p$,  
but increase above 1 as $p$ approaches 8.  
Near this value of $p$, $c_{E}$ 
reaches 1.81, while $c_{L}$ approaches 1.84.
As for eccentric orbits, we find that for large $p$ the energy and angular momentum approach the values given by the 
quadrupole approximation, but that for $p$ close to $6+2e$ they are better approximated by 
the energy and angular momentum radiated by a particle orbiting the black hole 
$N$ times on a circular orbit of radius $r_{p}|_{\rm min}=Mp/(1+e)$. 

\begin{figure}[!htb]
\vspace*{2.5in}
\special{hscale=31 vscale=31 hoffset=-13.0 voffset=190.0
         angle=-90.0 psfile=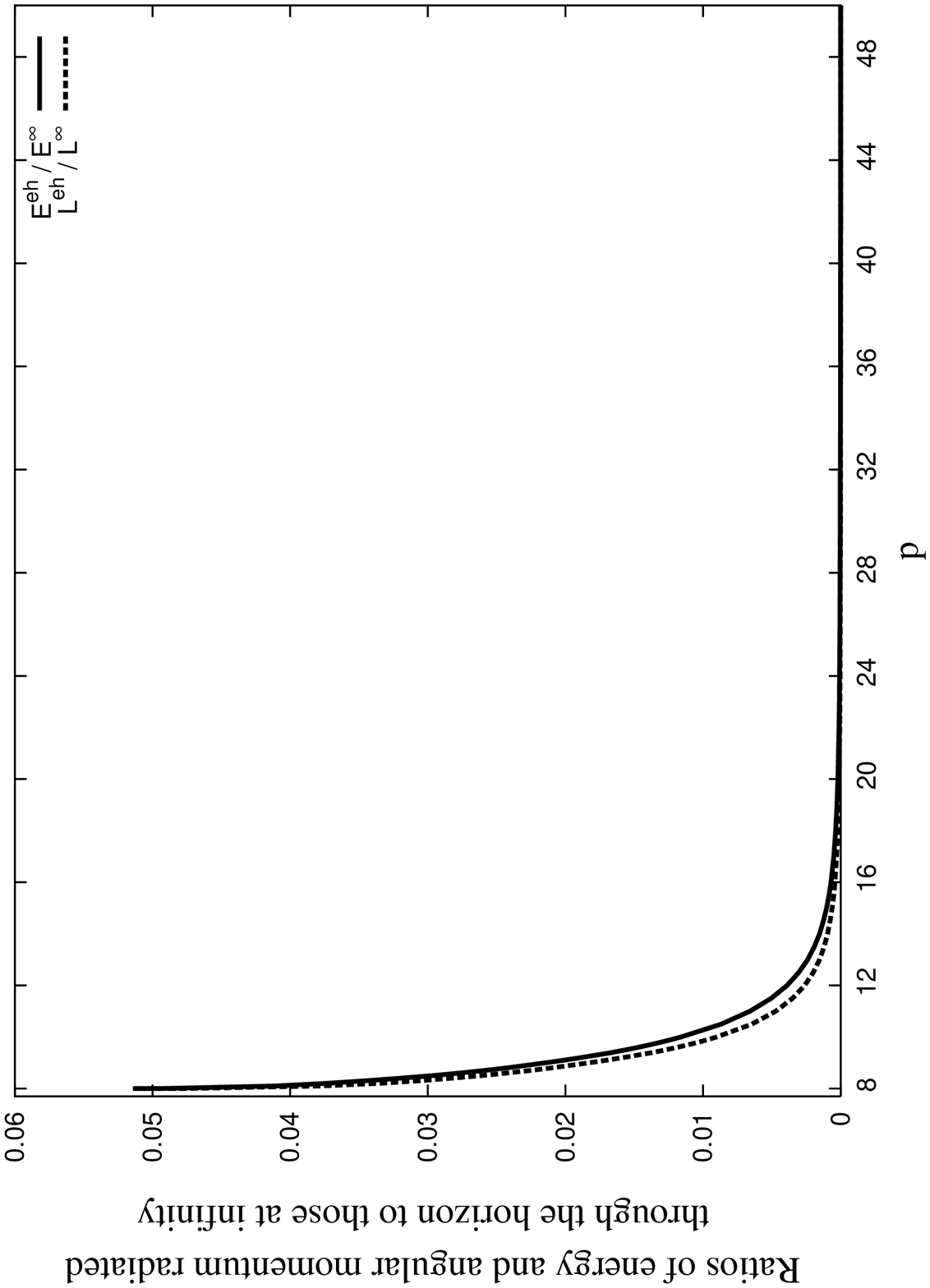}
\special{hscale=31 vscale=31 hoffset=225.0 voffset=190.0
         angle=-90.0 psfile=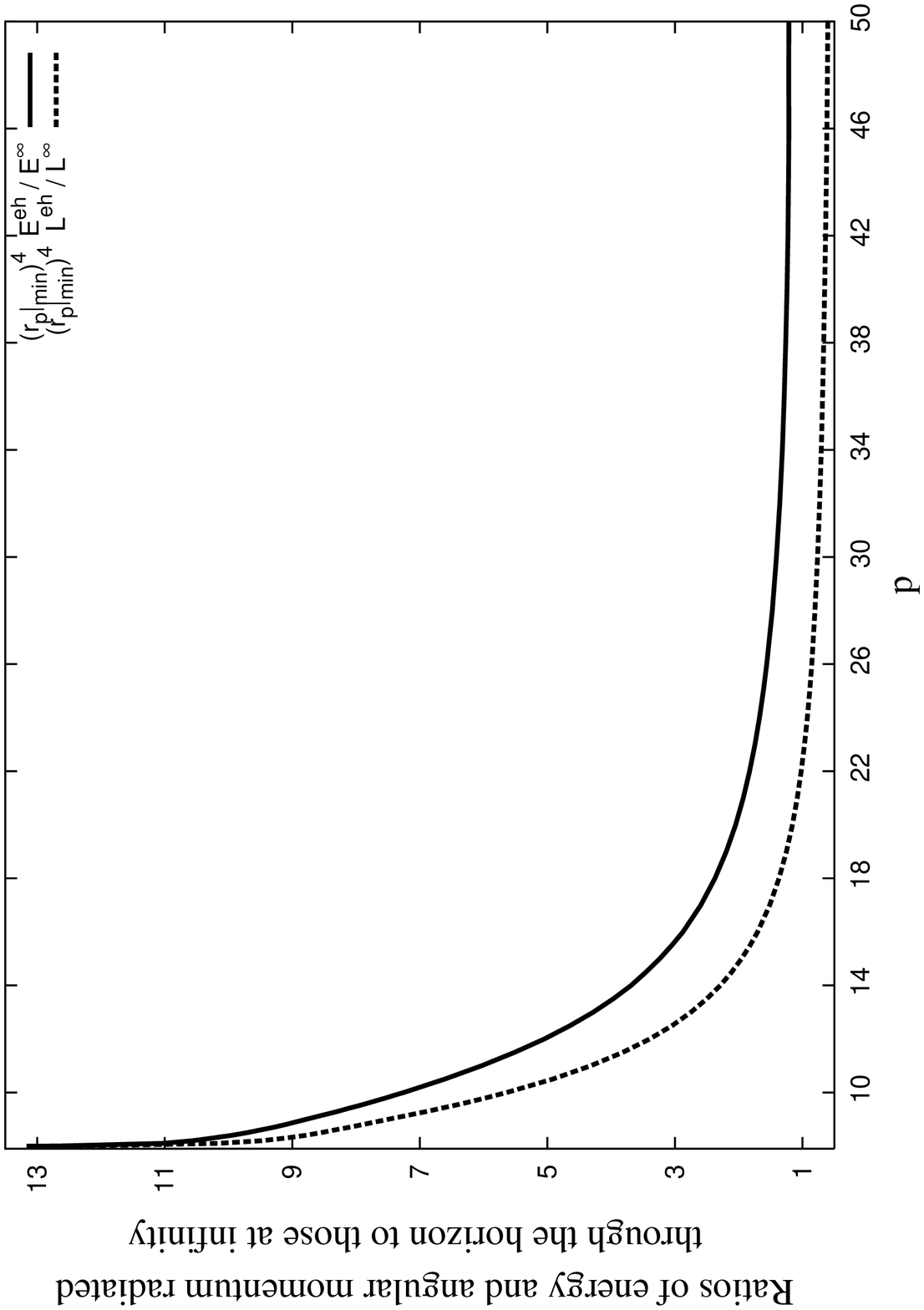}         
\caption{Black hole absorption for a particle following a parabolic geodesic.  The 
absorption of both energy and angular momentum is negligible until the particle 
reaches $p\approx 10$ or $r_{p}|_{\rm min}\approx 5M$.  
The right panel displays the same ratio normalized by $(M/r_{p}|_{\rm min})^{4}$, 
where $r_{p}|_{\rm min}$ is the radii at periastron.  Again, this factor is meaningful 
only for circular orbits, 
and is used only to illustrate the behaviour 
of black hole absorption as a function of $p$.}\label{fig:pEHflux}
\end{figure}
The argument given previously for eccentric orbits holds true for parabolic orbits: 
when $p\gtrsim 8$, black hole absorption is more important than for circular 
or eccentric orbit (see Fig.~\ref{fig:ecc_var}).  Our numerical results show that for $p> 10$, 
$E^{\rm eh}$ and $L^{\rm eh}$ account for less 
than $1\%$ of the total energy and angular momentum radiated, 
while for $p\lesssim 10$ they can contribute as much as  
$5\%$ of the total amounts (see Fig.~\ref{fig:pEHflux}).
Hence, for $p\lesssim 10$, 
black hole absorption contributes a few percents 
of the total energy and angular momentum radiated and needs to be included in an accurate computation. 
Black hole absorption is not determined as accurately as the energy and angular momentum 
radiated to infinity, but the error 
we make in evaluating it is never large enough to spoil our goal 
of $\sim 1\%$ overall accuracy.

\begin{table}[th!]
\caption{\label{tab:para}
Total energy and angular momentum radiated by a 
particle orbiting a Schwarzschild black hole in a parabolic orbit.  As usual, 
$E^{\infty}_{GR}$ and $L^{\infty}_{GR}$ denote the energy and angular momentum radiated to infinity, while 
$E^{\rm eh}_{GR}$ and $L^{\rm eh}_{GR}$ are the energy and angular momentum absorbed by the black hole.
The black hole absorption contributes to less than $1\%$ when $p\gtrsim 10$: for parabolic 
orbits with periastron smaller than $5M$, black hole absorption contributes a significant amount 
to the total energy and angular momentum radiated.
}
\begin{ruledtabular}
\begin{tabular}{rlccccc}
$p$ & $E^{\infty}_{GR}$ & $E^{\rm eh}_{GR}$ & $L^{\infty}_{GR}$ & $L^{\rm eh}_{GR}$ \\
\hline
8.00001&	3.6703E+00  &	1.8876E-01  &	3.0133E+01  &	1.5208E+00 \\
8.001  &	2.2809E+00  &	1.1260E-01  &	1.9088E+01  &	9.1166E-01 \\
8.201  &	7.1130E-01  &	2.6586E-02  &	6.6010E+00  &	2.2142E-01 \\
8.401  &	5.1740E-01  &	1.6534E-02  &	5.0433E+00  &	1.4244E-01 \\
8.601  &	4.0970E-01  &	1.1376E-02  &	4.1665E+00  &	1.0148E-01 \\
8.801  &	3.3767E-01  &	8.1988E-03  &	3.5706E+00  &	7.5175E-02 \\
9.0    &	2.8419E-01  &	6.0880E-03  &	3.1196E+00  &	5.7026E-02 \\
9.2    &	2.4409E-01  &	4.6044E-03  &	2.7756E+00  &	4.3891E-02 \\
9.4    &	2.1228E-01  &	3.5352E-03  &	2.4973E+00  &	3.4211E-02 \\
9.6    &	1.8644E-01  &	2.7473E-03  &	2.2665E+00  &	2.6951E-02 \\
9.8    &	1.6506E-01  &	2.1565E-03  &	2.0718E+00  &	2.1428E-02 \\
10.0   &	1.4712E-01  &	1.7072E-03  &	1.9048E+00  &	1.7176E-02 \\
10.5   &	1.1292E-01  &	9.8226E-04  &	1.5752E+00  &	1.0193E-02 \\
11.0   &	8.8979E-02  &	5.8626E-04  &	1.3320E+00  &	6.2819E-03 \\
11.5   &	7.1545E-02  &	3.6067E-04  &	1.1455E+00  &	3.9971E-03 \\
12.0   &	5.8467E-02  &	2.2778E-04  &	9.9827E-01  &	2.6152E-03 \\
12.5   &	4.8424E-02  &	1.4732E-04  &	8.7936E-01  &	1.7546E-03 \\
13.0   &	4.0567E-02  &	9.7321E-05  &	7.8158E-01  &	1.2036E-03 \\
13.5   &	3.4326E-02  &	6.5582E-05  &	7.0011E-01  &	8.4286E-04 \\
14.0   &	2.9303E-02  &	4.4999E-05  &	6.3136E-01  &	6.0123E-04 \\
14.5   &	2.5134E-02  &	3.1409E-05  &	5.7133E-01  &	4.3635E-04 \\
15.0   &	2.1774E-02  &	2.2273E-05  &	5.2088E-01  &	3.2171E-04 \\
16.0   &	1.6636E-02  &	1.1675E-05  &	4.3867E-01  &	1.8241E-04 \\
17.0   &	1.2978E-02  &	6.4484E-06  &	3.7499E-01  &	1.0869E-04 \\
18.0   &	1.0303E-02  &	3.7241E-06  &	3.2454E-01  &	6.7569E-05 \\
19.0   &	8.3029E-03  &	2.2354E-06  &	2.8383E-01  &	4.3565E-05 \\
20.0   &	6.7794E-03  &	1.3882E-06  &	2.5047E-01  &	2.8990E-05 \\
22.0   &	4.6735E-03  &	5.8355E-07  &	1.9949E-01  &	1.3895E-05 \\
24.0   &	3.3426E-03  &	2.6965E-07  &	1.6280E-01  &	7.2550E-06 \\
26.0   &	2.4638E-03  &	1.3447E-07  &	1.3549E-01  &	4.0539E-06 \\
28.0   &	1.8620E-03  &	7.1373E-08  &	1.1459E-01  &	2.3931E-06 \\
30.0   &	1.4374E-03  &	3.9949E-08  &	9.8213E-02  &	1.4781E-06 \\
32.0   &	1.1298E-03  &	2.3372E-08  &	8.5141E-02  &	9.4827E-07 \\
34.0   &	9.0223E-04  &	1.4201E-08  &	7.4534E-02  &	6.2832E-07 \\
36.0   &	7.2956E-04  &	8.9299E-09  &	6.5744E-02  &	4.2804E-07 \\
38.0   &	5.9799E-04  &	5.7886E-09  &	5.8484E-02  &	2.9873E-07 \\
40.0   &	4.9549E-04  &	3.8422E-09  &	5.2369E-02  &	2.1296E-07 \\
42.0   &	4.1455E-04  &	2.6163E-09  &	4.7168E-02  &	1.5469E-07 \\
44.0   &	3.4987E-04  &	1.8214E-09  &	4.2706E-02  &	1.1427E-07 \\
46.0   &	2.9763E-04  &	1.2887E-09  &	3.8849E-02  &	8.5683E-08 \\
48.0   &	2.5501E-04  &	9.3316E-10  &	3.5491E-02  &	6.5130E-08 \\
50.0   &	2.1993E-04  &	6.8343E-10  &	3.2550E-02  &	5.0118E-08 \\
\end{tabular}
\end{ruledtabular}
\end{table}
For completeness, in Table~\ref{tab:para} 
we display $E^{\infty,{\rm eh}}_{GR}$ and $L^{\infty,{\rm eh}}_{GR}$, the total energy and angular momentum 
radiated to 
infinity and through the event horizon, as returned by the time-domain code,
for a wide range of $p$ values.
Based on the accuracy obtained for circular and eccentric orbits, we estimate that the 
total energy and angular momentum lost to gravitational waves 
are calculated to a 
relative accuracy of $1\sim 2\%$. The actual accuracy is likely to be close to the accuracy achieved 
for circular orbits.  The reason for this is quite simple.  For parabolic orbits, 
there is no issue of performing a time-average, since the particle passes through periastron only once 
and we calculate the total energy for that motion.

%%%%%%%%%%%%%%%%%%%%%%%%%%%%%%%%%%%%%%%%%%%%%%%%%%%%%%%%%%%%%%%%%%%%%%%%%%%%
%%%%%%%%%%%%%%%%%%%%%%%%%%%%%%%%%%%%%%%%%%%%%%%%%%%%%%%%%%%%%%%%%%%%%%%%%%%%

\section{Conclusion} \label{sec:Conclusion}
The time-domain method can produce waveforms and compute 
the associated fluxes of energy and angular momentum to a relative accuracy of a 
few percents.  For circular orbits, the method is extremely reliable and 
produces fluxes with an overall accuracy of $1\%$ or better over the whole range of $p$ values explored.  
For eccentric orbits, the comparison 
with Cutler {\it et al.}~\cite{Cutler} 
is spoiled by the difficulty in performing a time average of the fluxes over a sufficiently long time. 
Because the disagreement arises from the differences in time averaging, 
the time-domain method is still capable of producing accurate waveforms for highly eccentric motion.  
We stress here that the limitation is in the computation of the time-averaged fluxes, not in 
obtaining the waveforms.
On the other hand, for geodesics with small 
eccentricities there is no such limitation and the time-domain results are in better agreement with those calculated by Cutler {\it et al.}~\cite{Cutler}.
In all cases, the time-domain method is capable of determining the fluxes accurately to $1\sim 2 \%$.
Similar accuracy is obtained for the total energy and angular momentum radiated by a particle 
traveling on a parabolic orbit.  

We also computed the absorption of energy and angular momentum by the black hole.
For circular orbits with $p>6$, this contribution can always be neglected, but not for 
orbits whose periastron is smaller than $5M$. For such orbits, black hole absorption  
contributes more than $1\% $ of the total fluxes and cannot be ignored.  
We showed that for $e=0.5$ it can constitute a correction as large as $2\%$ of the total fluxes; 
for parabolic orbits the contribution increases to $5\%$.

\section*{Acknowledgments}
This work was supported in part by the Ontario Graduate Scholarships and
by the Natural Sciences and Engineering Research Council of Canada.  
The author gratefully acknowledges the help of Eric Poisson and 
his comments on previous versions of this paper. 

%%%%%%%%%%%%%%%%%%%%%%%%%%%%%%%%%%%%%%%%%%%%%%%%%%%%%%%%%%%%%%%%%%%%%%%%%%%%
%%%%%%%%%%%%%%%%%%%%%%%%%%%%%%%%%%%%%%%%%%%%%%%%%%%%%%%%%%%%%%%%%%%%%%%%%%%%

\appendix
\section{A short review of black hole perturbation theory} \label{app:pt}
Since the pioneer work of T.~Regge and J.A.~Wheeler~\cite{RW} and F.J.~Zerilli~\cite{Zerilli}, 
perturbations of the Schwarzschild black hole have been 
studied extensively. 
Here we provide a short summary of the formalism, 
including the source terms appropriate for the 
Zerilli-Moncrief and Regge-Wheeler equations.

The perturbations of the Schwarzschild spacetime are described by a
linear perturbation tensor $h_{\mu \nu}=g_{\mu \nu}-g_{\mu \nu}^{Schwarzschild}$, 
where $g_{\mu \nu}$ is the metric of the perturbed spacetime and $g_{\mu \nu}^{Schwarzschild}$ 
the Schwarzschild solution.  
This tensor is written as a multipole expansion using scalar, vectorial, 
and tensorial spherical harmonics.  In Schwarzschild coordinates, we have
\begin{eqnarray}
h_{tt} &=& fH_{0}(t,r)\ Y^{lm}, \quad h_{tr}\ =\ H_{1}(t,r)\ Y^{lm}, \quad h_{rr} \ =\ fH_{2}(t,r)\ Y^{lm} \nonumber \\
h_{tA} &=& q_{0}(t,r)\ Z^{lm}_{A}+h_{0}(t,r)\ X^{lm}_{A}, \quad h_{rA}\ =\ q_{1}(t,r)\ Z^{lm}_{A}+h_{1}(t,r)\ X^{lm}_{A} \nonumber \\
h_{AB} &=& r^2\Bigl[K(t,r)\ U^{lm}_{AB} + G(t,r)\ V^{lm}_{AB}\Bigr] + h_{2}(t,r)\ W^{lm}_{AB},
\end{eqnarray}
where capital roman indices run over the angular coordinates ($\theta,\varphi$), $f=1-2M/r$, 
$Y^{lm}$ are the usual scalar harmonics, in term of which the vectorial and tensorial spherical harmonics are defined as
\[
Z^{lm}_{A} \ =\ Y^{lm}_{|A}, \quad X^{lm}_{A}\ =\ \varepsilon_{A}^{\ B}Y^{lm}_{|B}, \quad 
U^{lm}_{AB}\ =\ \Omega_{AB}Y^{lm}, \quad V^{lm}_{AB}\ = \ Y^{lm}_{|AB}+\frac{l(l+1)}{2}\Omega_{AB}Y^{lm}, \textrm{  and  } W^{lm}_{AB}\ = \ X^{lm}_{(A|B)}.
\]
Here, $\Omega_{AB}=(1,\sin^2\theta)$, a bar denotes the covariant derivative compatible with $\Omega_{AB}$, and $\varepsilon_{AB}$ is the Levi-Civita tensor on the unit two-sphere.
The coefficients of the multipole expansions have implicit $l$ and $m$ indices, and there is an implicit summation over 
these indices.  
Defined this way, $H_{0}$, $H_{1}$, $H_{2}$, $q_{0}$, $q_{1}$, $K$ and $G$ 
are even parity modes, 
while $h_{0}$, $h_{1}$ and $h_{2}$ are odd parity modes.  

By construction, the vectorial and tensorial spherical harmonics 
obey $S^{l,-m}=(-)^{m}S^{lm*}$, where $S$ is any spherical 
harmonic function.  (This relation holds for scalar spherical harmonics, 
and since vectorial and tensorial spherical harmonics are obtained 
by the action of {\em real} operators on $Y^{lm}$, it also applies to these functions.)  
An important consequence of this relation is that for real metric 
perturbations, the multipole moments must satisfy $M^{l,-m}=(-)^{m}M^{lm*}$, where 
$M^{lm}$ is any one of $H_{0}$, $H_{1}$, $H_{2}$, $q_{0}$, $q_{1}$, $K$, $G$, 
$h_{0}$, $h_{1}$ and $h_{2}$.
It is then easily established that $M^{lm}M^{*lm}=M^{l,-m}M^{*l,-m}$.  
This justifies folding the $m<0$ terms over to $m>0$ in Eqs.~(\ref{eqn:Etot}) and (\ref{eqn:Ltot}).

A gauge transformation can be used to eliminate four of the metric 
perturbations.  In the Regge-Wheeler gauge, this freedom is used to set 
$q_{0}=q_{1}=G=0$, and $h_{2}=0$. 
The two scalar fields 
\begin{eqnarray}
\psi_{\rm ZM} &=& \frac{r}{\lambda + 1} \Biggl[ K + \frac{f}{\Lambda}
\biggl( H_2 - r \frac{\partial}{\partial r} K \biggr) \Biggr], \label{eqn:ZMfun} \\
\psi_{\rm RW} &=& -\frac{f}{r}\, h_1, \label{eqn:RWfun}
\end{eqnarray}
where $\lambda=(l+2)(l-1)/2$ and $\Lambda =\lambda+ 3M/r$, are the Zerilli-Moncrief and Regge-Wheeler functions, respectively. 
Their evolution is governed by Eq.~(\ref{eqn:wave}) with the potentials
\begin{eqnarray} \label{eqn:pot}
V_{\rm ZM} &=& \frac{f}{r^2\Lambda^2}\Bigg[2\lambda^2\left(\lambda+1+\frac{3M}{r}\right)+\frac{18M^2}{r^2}\left(\lambda+\frac{M}{r}\right)\Bigg], \nonumber \\
V_{\rm RW} &=& \frac{f}{r^2}\Bigg[l(l+1)-\frac{6M}{r}\Bigg],
\end{eqnarray}
and the source terms 
\begin{eqnarray*}
S_{\rm ZM} &=& \frac{1}{(\lambda+1)\Lambda}\, \Biggl\{ 
	    r^2 f \biggl( f^2 \frac{\partial}{\partial r} Q^{tt} - \frac{\partial}{\partial r} Q^{rr} \biggr) 
	    + r(\Lambda - f) Q^{rr} + rf^2 Q^\flat \nonumber \\
	   & & - \frac{f^2}{r\Lambda} \Bigl[ \lambda(\lambda-1)r^2 + (4\lambda-9)Mr + 15M^2
	    \Bigr] Q^{tt} \Biggr\} + \frac{2f}{\Lambda}\, Q^{r} - \frac{f}{r}\, Q^\sharp \\
S_{\rm RW} &=& \frac{f}{r}\, \Biggl[ \frac{2}{r} \biggl( 1 - \frac{3M}{r} \biggr) P - f \frac{\partial}{\partial r} P + P^r \Biggr].
\end{eqnarray*}
These are constructed from the perturbing stress-energy tensor $T^{\mu \nu}$ and we have defined    
\[ 
Q^{ab} = 8\pi \int T^{ab} Y^{lm*}\, d\Omega, \qquad
Q^a = \frac{16\pi r^2}{l(l+1)}\, \int T^{aA} Z^{lm*}_A\, d\Omega,
\]
\vspace*{-10pt}
\[
Q^\flat = 8\pi r^2 \int T^{AB} U^{lm*}_{AB}\, d\Omega, \qquad
Q^\sharp = \frac{32\pi r^4}{(l-1)l(l+1)(l+2)}\, \int T^{AB}
V^{lm*}_{AB}\, d\Omega, 
\]
and  
\[ 
P^a = \frac{16\pi r^2}{l(l+1)}\, \int T^{aA} X^{lm*}_{A}\, d\Omega,
\qquad  
P = \frac{16 \pi r^4}{(l-1)l(l+1)(l+2)}\, \int T^{AB} W^{lm*}_{AB}\,
d\Omega; 
\]
lower-case roman indices run over $t$ and $r$, the integration is over the unit 
two-sphere with $d\Omega = \sin\theta\, d\theta d\varphi$, 
and $T^{ab}$, $T^{aA}$ and $T^{AB}$ are components 
of $T^{\mu \nu}$.  For a particle traveling on 
a geodesic with proper time $\tau$, coordinates $z^{\mu}_{p}(\tau)$, 
and four-velocity $u^{\mu}(\tau)$, the 
stress-energy tensor is
\begin{eqnarray}
T^{\mu\nu} &=& \mu \int d\tau\ (-g)^{-1/2}\  u^{\mu}u^{\nu}\ \delta^{4}\left[x^{\alpha}-z^{\alpha}_{p}\right],
\end{eqnarray}
where $\delta^{4}[x^{\alpha}-z^{\alpha}]$ is a four-dimensional Dirac functional.

These expressions for the source terms can be used, combined with the geodesic equations of the Schwarzschild spacetime, to calculate explicit expressions for the 
factors $G(t,r)$ and $F(t,r)$ appearing in Eq.~(\ref{eqn:src}).  
For even parity modes (ZM), we get
\begin{eqnarray} \label{eqn:SZM}
G(r,t)&=& a \ Y^{*}(t) + b \ Z^{*}_{\varphi}(t) + c \ U^{*}_{\varphi \varphi}(t)+ d \ V^{*}_{\varphi \varphi}(t) \nonumber\\
F(r,t)&=& \frac{8\pi}{\lambda+1}\frac{f^2}{\Lambda}\frac{\tilde{V}^2}{\tilde{E}}\ Y^{*}(t),
\end{eqnarray}
where $\tilde{V}^2=f\left(1+\tilde{L}^2/r^2\right)$, 
\[
a = \frac{8\pi}{\lambda+1}\frac{f^2}{r\Lambda^2}\Biggl\{ \frac{6M}{r}\tilde{E}-\frac{\Lambda}{\tilde{E}}\biggl[\lambda+1-\frac{3M}{r}+\frac{\tilde{L}^2}{r^2}\left(\lambda+3-\frac{7M}{r}\right)\biggr]\Biggr\},
\]
and 
\[
b = \frac{16 \pi}{\lambda+1} \frac{\tilde{L}}{\tilde{E}}\frac{f^2}{r^2\Lambda}\ u^{r}, \qquad
c = \frac{8\pi}{\lambda+1} \frac{\tilde{L}^2}{\tilde{E}}\frac{f^3}{r^3 \Lambda}, \qquad
d = -32 \pi \frac{(l-2)!}{(l+2)!} \frac{\tilde{L}^2}{\tilde{E}}\frac{f^2}{r^3}.
\]
Finally, the source terms for odd parity modes (RW) are
\begin{eqnarray} \label{eqn:SRW}
G(r,t) &=&\frac{f^{2}}{r^3}\Biggl[\frac{4}{r}\left(1-\frac{3M}{r}\right)\alpha +\beta\Biggr], \nonumber \\
F(r,t) &=& -\alpha\frac{f^3}{r^3}, 
\end{eqnarray}
where
\begin{eqnarray*}
\alpha &=&16 \pi \frac{(l-2)!}{(l+2)!} \frac{\tilde{L}^2}{\tilde{E}}\ W^{*lm}_{\varphi \varphi}(t), \quad \beta\ =\ \frac{8 \pi}{\lambda+1} \frac{\tilde{L}}{\tilde{E}}\ u^{r}\ X^{*lm}_{\varphi}(t).
\end{eqnarray*}

Note that $G(r,t)$ and $F(r,t)$ contain 
scalar, vectorial and tensorial harmonic functions evaluated 
at the angular position of the particle $\varphi_{p}(t)$. 
For example, for even modes, some terms in $G$ and $F$ are proportional to 
$Y^{lm}(\pi/2,\varphi_{p}(t))$. Because the orbital motion takes place in the 
equatorial plane, each spherical harmonic function is evaluated at $\theta_{p}=\pi/2$.
A useful consequence of this is that the source term for the 
Zerilli-Moncrief function vanishes 
when $l+m$ is odd, while the source term for the Regge-Wheeler function 
vanishes when $l+m$ is even. This was used in Eq.~(\ref{eqn:eflux}) and Eq~(\ref{eqn:lflux}).

Once $\psi_{\rm ZM}$ and $\psi_{\rm RW}$ are found by solving Eq.~(\ref{eqn:wave}),  
with the source terms of Eqs.~(\ref{eqn:SZM}) and (\ref{eqn:SRW}), 
the perturbation tensor can be reconstructed. In the Regge-Wheeler gauge, 
we have
\begin{eqnarray} \label{eqn:MetricEven}
K     &=& f \frac{\partial}{\partial r}\psi_{\rm ZM} +A(r)\psi_{\rm ZM}-\frac{r^2f^2}{(\lambda+1)\Lambda}Q^{tt}, \nonumber \\
H_{2} &=& \frac{\Lambda}{f}\Bigg[\frac{\lambda+1}{r}\psi_{\rm ZM}-K\Bigg] +r\frac{\partial}{\partial r}K, \nonumber \\
H_{1} &=& r\frac{\partial}{\partial t}\Bigg[\frac{\partial}{\partial r}\psi_{\rm ZM}+B(r)\psi_{\rm ZM}\Bigg]-\frac{r^2}{\lambda+1}\Bigg[Q^{tr}+\frac{rf}{\Lambda}\frac{\partial}{\partial t}Q^{tt}\Bigg], \nonumber \\
H_{0} &=& H_{2}+Q^{\sharp},
\end{eqnarray}
for even parity modes, where we have defined $A(r)=[\lambda(\lambda+1)+3M/r\left(\lambda+2M/r\right)]/(r\Lambda)$ and 
$B(r)=[\lambda(1-3M/r)-3M^2/r^2]/(rf\Lambda)$.  
For odd parity modes, the reconstructed metric perturbations are  
\begin{eqnarray} \label{eqn:MetricOdd}
h_{0} &=& -f \int_{-\infty}^{t} dt'\Bigg[\frac{\partial}{\partial r}\bigg(r\psi_{\rm RW}(t',r)\bigg)+P\Bigg], \nonumber \\
h_{1} &=& -rf^{-1}\psi_{\rm RW}.
\end{eqnarray}

%%%%%%%%%%%%%%%%%%%%%%%%%%%%%%%%%%%%%%%%%%%%%%%%%%%%%%%%%%%%%%%%%%%%%%%%%%%%
%%%%%%%%%%%%%%%%%%%%%%%%%%%%%%%%%%%%%%%%%%%%%%%%%%%%%%%%%%%%%%%%%%%%%%%%%%%%

\section{Radiation Zone Fluxes and Black Hole Absorption} \label{app:Fluxes}
The fluxes of energy and angular momentum can be obtained from Isaacson's 
stress-energy tensor for gravitational waves:
\begin{eqnarray} \label{eqn:GWtensor}
T^{GW}_{\mu \nu}=\frac{1}{64\pi} \left<h^{\alpha \beta}_{\ \ ;\mu} h_{\alpha \beta;\nu}\right>,
\end{eqnarray}
where $\left<\ldots\right>$ denotes an average over a region of spacetime large compared 
with the wavelength of the radiation.  Typically, $T_{\mu\nu}$ 
can be defined when the wavelength of the radiation, $\lambda$, is 
small compared to a typical radius of curvature ${\mathcal R}$.  
By definition, $\lambda \ll {\mathcal R}$ in the radiation zone and the stress-energy 
tensor for gravitational waves can be defined there.  There is a second region where the condition 
$\lambda \ll {\mathcal R}$ is satisfied: 
A stationary observer near $r=2M$ sees ${\mathcal R}\sim 2M$, but  
the radiation is strongly blueshifted and $\lambda\rightarrow 0$;  that this 
is the case is clear from the divergence in Eq.~(\ref{eqn:Tblue}) below.

Because the Schwarzschild black hole is static and axially symmetric, 
it possesses two Killing vectors,
$\,_{(t)}\xi^{\alpha}$ and $\,_{(\varphi)}\xi^{\alpha}$,
that can be used, in conjunction with $T_{\mu\nu}$, 
to obtain expressions for the fluxes of energy and angular momentum 
through a surface $\Sigma$ at constant $r$.  We have that (dropping the ``$GW$'' on the stress-energy tensor)
\begin{eqnarray}
dE = -\int_{\Sigma} T^{\mu}_{\ \nu}\ \,_{(t)}\xi^{\nu}   \  d\Sigma_{\mu}\\
dL =  \int_{\Sigma} T^{\mu}_{\ \nu}\ \,_{(\varphi)}\xi^{\nu}\  d\Sigma_{\mu}, 
\end{eqnarray}
where $d \Sigma_{\mu}$ is an outward oriented surface element on $\Sigma$.  In Schwarzschild 
coordinates, these expressions reduce to 
\begin{eqnarray} \label{eqn:FluxGeneral}
\dot{E} &=& -\epsilon r^2f \int d\Omega\ T_{tr}\ , \nonumber \\
\dot{L} &=& \epsilon r^2f \int d\Omega\ T_{r \varphi}\ ,
\end{eqnarray}
where an overdot indicates differentiation with respect to $t$, 
and $\epsilon$ is 1 when calculating the fluxes in the radiation zone and -1 when calculating 
black hole absorption.  Because the event horizon is a null surface, 
it is conceptually better to use $dE/dv$ and $dL/dv$ in place of $dE/dt$ and $dL/dt$ (with $v=t+r^{*}$). 
Similarly, because radiation travels to ${\mathcal J}^{+}$, which is another 
null surface, it is conceptually better to use $dE/du$ and $dL/du$ instead of $dE/dt$ and $dL/dt$ for outgoing fluxes (with $u=t-r^{*}$).  
However, because we numerically extract the gravitational waveforms 
at finite values of $r^{*}$, without ever reaching the event horizon or ${\mathcal J}^{+}$, 
the use of $dE/dt$ and $dL/dt$ is a better representation of our numerical procedure.  

We provided a summary of the perturbation formalism in Appendix~\ref{app:pt}.
This included explicit formulae to reconstruct the metric perturbations 
in the Regge-Wheeler gauge.  To construct $T_{tr}$ and $T_{r\varphi}$, 
it proves convenient to extract the radiative part of the perturbation tensor. 
To this end, we introduce the null tetrad $l^{\mu}$, $n^{\mu}$, and $m^{\mu}$: $l^{\mu}$ 
and $n^{\nu}$ are null vectors tangent 
to outgoing and ingoing rays, respectively, while $m^{\mu}$ is a complex 
null vector (with complex conjugate $\bar{m}^{\mu}$) on the two-sphere.  
They satisfy the relations 
\begin{eqnarray} \label{eqn:tetradConditions}
l^{\mu}l_{\mu} = 0  =  n^{\mu}n_{\mu},\qquad m^{\mu}m_{\mu} = 0  =  \bar{m}^{\mu}\bar{m}_{\mu}, \qquad l^{\mu}n_{\mu} = -1 =  -m^{\mu}\bar{m}_{\mu}. 
\end{eqnarray}  
In Schwarzschild coordinates, their components are 
\begin{eqnarray} \label{eqn:tetrad}
l_{\mu} = (-1,f^{-1},0,0), \qquad n_{\mu}  = -\frac{1}{2}(f,1,0,0), \qquad m_{\mu} = \frac{\sqrt{2}}{2}r(0,0,1,\imath \sin\theta).
\end{eqnarray}

To find the perturbation tensor in a radiation gauge, 
we seek a gauge vector $\xi^{\mu}$ that generates the 
transformation between the Regge-Wheeler gauge and the radiation gauge.
Under such an infinitesimal coordinate transformation, 
the perturbation field transforms as $h^{RG}_{\mu\nu}=h^{RW}_{\mu\nu}- \xi_{(\mu;\nu)}$, 
where $RG$ stands for radiation gauge and $RW$ for Regge-Wheeler gauge.
Using the spherical harmonic functions introduced previously, 
the vector $\xi^{\mu}$ can be expressed as a multipole expansion~\cite{RW}:
\begin{eqnarray}
\xi_{\mu}^{(even)} &=& (\alpha_{t}\ Y^{lm},\alpha_{r}\ Y^{lm},r^2\beta \ Z^{lm}_{A}),\nonumber \\
\xi_{\mu}^{(odd)}  &=& (0,0, r^2\kappa \ X^{lm}_{A}),
\end{eqnarray}
where $\alpha_{t}$, $\alpha_{r}$, $\beta$ and $\kappa$ are freely specifiable gauge functions. 
Combining this with the multipole expansion for the perturbation tensor in 
the Regge-Wheeler gauge, we obtain the transformation law for the each perturbation mode:
\begin{eqnarray} \label{eqn:EGSch}
H_{0}^{RG} &=& H^{\rm RW}_{0} - 2f^{-1}\Bigg(\dot{\alpha}_{t}-\frac{M}{r^2}f\alpha_{r}\Bigg), \nonumber \\
H_{1}^{RG} &=& H^{\rm RW}_{1} - \Bigg(\alpha_{t}{'}+\dot{\alpha}_{r}-\frac{2M}{r^2}f^{-1}\alpha_{t}\Bigg), \nonumber \\
H_{2}^{RG} &=& H^{\rm RW}_{2} - 2f\Bigg(\alpha_{r}{'}+\frac{M}{r^2}f^{-1}\alpha_{r}\Bigg), \nonumber \\
q_{t}^{RG} &=& - \Bigg(\alpha_{t}+r^2\dot{\beta}\Bigg), \nonumber \\
q_{r}^{RG} &=& - \Bigg(\alpha_{r}+r^2\beta'\Bigg), \nonumber \\
K^{RG}	   &=& K^{\rm RW} -\Bigg(\frac{2f}{r}\alpha_{r} -l(l+1)\beta\Bigg), \nonumber \\
G^{RG}	   &=& -2\beta,
\end{eqnarray}
for the transformation of even parity modes, and 
\begin{eqnarray} \label{eqn:OGSch}
h^{RG}_{t} &=& h^{\rm RW}_{0}-r^2\dot{\kappa}, \nonumber \\
h^{RG}_{r} &=& h^{\rm RW}_{1}-r^2\kappa{'}, \nonumber \\
h^{RG}_{2} &=& -2r^2\kappa
\end{eqnarray}
for odd modes;  an overdot designates a $t$-derivative and a prime an $r$-derivative.
In the next two sections, we develop solutions appropriate to 
outgoing and ingoing radiation gauges, respectively.

%%%%%%%%%%%%%%%%%%%%%%%%%%%%%%%%%%%%%%%%%%%%%%%%%%%%%%%%%%%%%%%%%%%%%%%%%%%%
%%%%%%%%%%%%%%%%%%%%%%%%%%%%%%%%%%%%%%%%%%%%%%%%%%%%%%%%%%%%%%%%%%%%%%%%%%%%

\subsection{Outgoing radiation gauge} 
In an outgoing radiation gauge $h_{\mu\nu}^{ORG}$, the perturbation 
tensor satisfies~\cite{Chrzanowsky}
\begin{eqnarray} \label{eqn:ORGconditions}
h^{ORG}_{\mu\nu}n^{\mu}n^{\nu} &=& 0, \nonumber \\
h^{ORG}_{\mu\nu}n^{\mu}m^{\nu} &=& 0 \ =\ h^{ORG}_{\mu\nu}n^{\mu}\bar{m}^{\nu}, \nonumber \\
h^{ORG}_{\mu\nu}n^{\mu}l^{\nu} &=& 0 \ =\ h^{ORG}_{\mu\nu}m^{\mu}\bar{m}^{\nu}.
\end{eqnarray}
The first two conditions indicate the $h^{ORG}_{\mu \nu}$ is transverse to outgoing null rays, 
while the last condition indicates that it is traceless.  
Eqs.~(\ref{eqn:ORGconditions}) involve five conditions, one too many for 
the specification of a gauge.  But this system is not overdetermined: once four of these 
equations are enforced, the fifth is found to be satisfied automatically in the radiation zone.

From Eqs.~(\ref{eqn:EGSch}) and (\ref{eqn:OGSch}), 
the gauge conditions can be expressed in terms of multipole moments.  To leading-order in $r^{-1}$, we get 
\begin{eqnarray} \label{eqn:eorg}
4\left(\dot{\alpha}_{t}-\dot{\alpha}_{r}\right)&=&H^{\rm RW}_{0}+H^{\rm RW}_{2}-2H^{\rm RW}_{1},\nonumber \\
\alpha_{t}-\alpha_{r}+2r^2\dot{\beta}&=&0, \nonumber \\
\dot{\alpha}_{t}+\dot{\alpha}_{r} &=& 0, \nonumber \\
\frac{2}{r}\alpha_{r}&=& -K^{\rm RW},
\end{eqnarray}
for even parity modes, and
\begin{eqnarray} \label{eqn:oorg}
2r^2\dot{\kappa} &=& h^{\rm RW}_{1}-h^{\rm RW}_{0}, 
\end{eqnarray}
for odd parity modes.  
We used $\partial/\partial t = - \partial/\partial r +{\mathcal O}(r^{-1})$, appropriate in the radiation zone, 
to eliminate $r$-derivatives in favor of $t$-derivatives.  

The right-hand side of these equations 
contains the perturbation tensor in the Regge-Wheeler gauge.  
From Eq.~(\ref{eqn:MetricEven}), 
we find that, in the radiation zone, even parity modes 
have the asymptotic form 
\[
K^{\rm RW} \approx -\dot{\psi}_{\rm ZM}, \quad H^{\rm RW}_{2}\approx H^{\rm RW}_{0}\approx-H^{\rm RW}_{1}\approx r\ddot{\psi}_{\rm ZM}, \nonumber
\]
while, from Eq.~(\ref{eqn:MetricOdd}), we find that odd parity modes are given asymptotically by 
\[
h_{0} \approx -h_{1} \approx r\psi_{\rm RW}. \nonumber
\]

Solutions to the gauge transformation are then easy to find.
For even modes, 
the last of Eq.~(\ref{eqn:eorg}) yields $\alpha_{r}=- (1/2) r\dot{\psi}_{\rm ZM}$.  
The other gauge functions are easily obtained from the remaining equations of 
Eq.~(\ref{eqn:eorg}): $\alpha_{t}=-\alpha_{r}$ and $\beta = -\psi_{\rm ZM}/(2r)$.
For odd modes, direct integration of Eq.~(\ref{eqn:oorg}) yields 
$\kappa =1/r \int_{-\infty}^{t} d t'\psi_{\rm RW}(t')$. 
Going back to Eqs.~(\ref{eqn:EGSch}) and (\ref{eqn:OGSch}), 
we can reconstruct the perturbation tensor in the radiation zone.  
Solving these for $G^{ORG}$ and $h_{2}^{ORG}$, we get 
\begin{eqnarray} \label{eqn:hORG}
h_{AB}^{ORG} = r\left(\psi_{\rm ZM}(t)V^{lm}_{AB}-2 \int_{-\infty}^{t}dt'\psi_{\rm RW}(t')W_{AB}^{lm}\right) +{\mathcal O}(1).
\end{eqnarray}
In this expression, the Zerilli-Moncrief and Regge-Wheeler functions 
have $l$ and $m$ multipole 
indices and there is an implicit summation over them.  Dropping the ${\mathcal O}(1)$ 
term, we refer to $h_{AB}^{ORG}$ as the radiative part of the perturbation tensor 
in the radiation zone, because it contains all of the information about energy and 
angular momentum carried to infinity.
This last expression can 
be re-written in terms of the two gravitational-wave polarizations, $h_{+}=h^{ORG}_{\theta \theta}/r^2$,  
and $h_{\times}=h^{ORG}_{\theta \varphi}/(r^2\sin\theta)$.  The result is Eq.~(\ref{eqn:hphcO}).

%%%%%%%%%%%%%%%%%%%%%%%%%%%%%%%%%%%%%%%%%%%%%%%%%%%%%%%%%%%%%%%%%%%%%%%%%%%%
%%%%%%%%%%%%%%%%%%%%%%%%%%%%%%%%%%%%%%%%%%%%%%%%%%%%%%%%%%%%%%%%%%%%%%%%%%%%

\subsection{Ingoing radiation gauge}
To obtain the radiative part of the gravitational field 
in the vicinity of the event horizon, we impose an ingoing 
radiation gauge.  We seek a solution to leading order in $f\rightarrow 0$,
the expansion parameter near the horizon.

The ingoing radiation gauge can be obtained from Eq.~(\ref{eqn:ORGconditions}), 
by making the replacement $l^{\mu}\leftrightarrow n^{\mu}$ of the tetrad vectors. 
The same comment about the number of gauge conditions can be made here: 
only four gauge conditions need to be imposed, and the fifth condition 
is then satisfied automatically.

From Eq.~(\ref{eqn:EGSch}), Eq.~(\ref{eqn:OGSch}), 
and the ingoing radiation gauge conditions, 
we obtain 
\begin{eqnarray} \label{eqn:eirg}
4(\dot{\alpha}_{t}+\dot{\alpha}_{r})-\frac{1}{M}\left(\alpha_{t}+\alpha_{r}\right) &=& 2f\left(H^{\rm RW}_{2}+H^{\rm RW}_{1}\right), \nonumber \\
\dot{\alpha}_{t}-\dot{\alpha}_{r} &=& 0, \nonumber \\
8M^2\dot{\beta}+\alpha_{t}+\alpha_{r} &=& 0, \nonumber \\
\frac{1}{M}\alpha_{r}-\beta &=& K^{\rm RW},
\end{eqnarray}
for the gauge transformation of even parity modes, and
\begin{eqnarray} \label{eqn:oirg}
r^2\dot{\kappa}=r(h^{\rm RW}_{0}+fh^{\rm RW}_{1}),
\end{eqnarray}
for odd parity modes: we substituted $\alpha_{r}\rightarrow f^{-1}\alpha_{r}$ 
in Eq.~(\ref{eqn:EGSch}), we used $\partial/\partial r^{*} = \partial/\partial t + {\mathcal O}(f)$ to 
eliminate derivatives with respect to $r^{*}$, and 
an overdot denotes a time derivative.

The asymptotic form of the metric perturbations in the Regge-Wheeler gauge is obtained from Eqs.~(\ref{eqn:MetricEven}) 
and (\ref{eqn:MetricOdd}).  We get 
\[
K^{\rm RW} \approx \dot{\psi}_{\rm ZM} +\frac{\lambda+1}{2M}\psi_{\rm ZM}, \quad H^{\rm RW}_{2}=H^{\rm RW}_{0}=H^{\rm RW}_{1}=f^{-1}\left(2M\ddot{\psi}_{\rm ZM}-\frac{1}{2}\dot{\psi}_{\rm ZM}\right), \nonumber 
\]
for even parity modes, and 
\[
h^{\rm RW}_{0}=fh^{\rm RW}_{1}=-2M\psi_{\rm RW}, \nonumber 
\]
for odd parity ones.  These can be inserted back into 
Eqs.~(\ref{eqn:eirg}) and (\ref{eqn:oirg}).  The solution to the gauge transformation then 
proceeds as follow. 
For even parity modes, the second and third equations 
yield $\alpha_{r}=\alpha_{t}$ and $4M^2\dot{\beta}=-\alpha_{t}$.  
These can be substituted into the first and the time derivative of the fourth 
of Eq.~(\ref{eqn:eirg}), to yield a system of equation for $\alpha_{t}$: 
\begin{eqnarray}
\dot{\alpha}_{t}-\frac{1}{4M}\alpha_{t} &=& M\ddot{\psi}_{\rm ZM}-\frac{1}{4}\dot{\psi}_{\rm ZM}, \\
\dot{\alpha}_{t}-\frac{\lambda+1}{2M}\alpha_{t} &=& M\ddot{\psi}_{\rm ZM}+\frac{\lambda+1}{2}\dot{\psi}_{\rm ZM}.
\end{eqnarray}
Eliminating the time derivative by subtraction, we find $\alpha_{t}=M\dot{\psi}_{\rm ZM}$. 
For odd parity modes, integration of Eq.~(\ref{eqn:oirg}), 
combined with the asymptotic form of $h^{\rm RW}_{0}$ and $h^{\rm RW}_{1}$, yields $\kappa =- 1/(2M)\int dt' \psi_{\rm RW}(t')$.

From these, and Eqs.~(\ref{eqn:EGSch}) and (\ref{eqn:OGSch}), 
we obtain $G^{IRG}$ and $h_{2}^{IRG}$, which are used to 
reconstruct the gravitational perturbation tensor:
\begin{eqnarray} \label{eqn:hIRG}
h^{IRG}_{AB} = 2M\Bigg[\psi_{\rm ZM}V^{lm}_{AB}+2\int dt' \psi_{\rm RW}(t')W^{lm}_{AB}\Bigg] + {\mathcal O}(f).
\end{eqnarray}
Again, the Zerilli-Moncrief and Regge-Wheeler functions 
have $l$ and $m$ multipole indices, 
and there is an implicit summation over them.
The components $h^{IRG}_{AB}$ (without the ${\mathcal O}(f)$ correction) 
contains all the information about the energy and angular momentum absorbed by the black hole. 
It is then meaningful to refer to these components as the radiative part 
of the perturbation tensor in the vicinity of the event horizon.
In analogy with the far zone definitions, 
the two gravitational-wave polarizations are defined as 
$h_{+}=h^{IRG}_{\theta\theta}/4M^2$ 
and $h_{\times}=h^{IRG}_{\theta\varphi}/(4M^2\sin\theta)$.  
They are given in Eq.~(\ref{eqn:hphcI}).  

%%%%%%%%%%%%%%%%%%%%%%%%%%%%%%%%%%%%%%%%%%%%%%%%%%%%%%%%%%%%%%%%%%%%%%%%%%%%
%%%%%%%%%%%%%%%%%%%%%%%%%%%%%%%%%%%%%%%%%%%%%%%%%%%%%%%%%%%%%%%%%%%%%%%%%%%%

\subsection{Radiation zone fluxes} \label{app:org}
In terms of its tetrad components, 
the perturbation tensor in the outgoing radiation gauge is 
\begin{eqnarray} \label{eqn:ORGhtetrad}
h^{ORG}_{\mu \nu} &=& h_{ll}\ n_{\mu}n_{\nu} + 2\left(h_{l\bar{m}}n_{(\mu}m_{\nu)}+h_{lm}n_{(\mu}\bar{m}_{\nu)}\right) \nonumber \\
		  &+& h_{\bar{m}\bar{m}}m_{\mu}m_{\nu}+h_{mm}\bar{m}_{\mu}\bar{m}_{\nu},
\end{eqnarray}
where $h_{vv}=h_{\mu\nu}v^{\mu}v^{\nu}$, for any vector $v^{\mu}$ belonging to the tetrad.
To leading-order in $r^{-1}$, the tetrad components are 
\begin{eqnarray} \label{eqn:ORGrTetrad}
h_{ll} &\sim& {\mathcal O}(r^{-2}), \nonumber \\
h_{lm} &\sim& {\mathcal O}(r^{-2}), \nonumber \\
h_{\bar{m}\bar{m}} &=& \frac{1}{r^2}h^{ORG}_{AB}\ \bar{m}^{A}\bar{m}^{B}, \nonumber \\
h_{mm} &=& \frac{1}{r^2}h^{ORG}_{AB}\ m^{A}m^{B},
\end{eqnarray}
where the vector $m^{A}=r\ (\partial \theta^A/\partial x^{\mu})\ m^{\mu}$, 
and $h^{ORG}_{AB}$ is given in Eq.~(\ref{eqn:hORG}).

Calculating the covariant derivative of $h^{ORG}_{\mu\nu}$ and 
substituting the result in Eq.~(\ref{eqn:GWtensor}), we get 
\begin{eqnarray} \label{eqn:Ttmp}
T_{\mu \nu} &=& \frac{1}{64\pi}\Bigg[\eta^{\alpha \beta}_{\ \ \mu}\eta^{*}_{\alpha \beta \nu}+\eta^{\alpha \beta}_{\ \ \mu}\rho^{*}_{\alpha \beta \nu}+ \rho^{\alpha \beta}_{\ \ \mu}\eta^{*}_{\alpha \beta \nu}+\rho^{\alpha \beta}_{\ \ \mu}\rho^{*}_{\alpha \beta \nu}\Bigg] +c.c.\ , 
\end{eqnarray}
where 
\begin{eqnarray} \label{eqn:etarho_ORG}
\eta_{\alpha \beta \mu} &=& h_{ll,\mu}n_{\alpha}n_{\beta} + 2 h_{lm,\mu}n_{(\alpha}\bar{m}_{\beta)}+ 2 h_{l\bar{m},\mu}n_{(\alpha}m_{\beta)} \nonumber \\
			&+& h_{mm,\mu}\bar{m}_{\alpha}\bar{m}_{\beta} +h_{\bar{m}\bar{m},\mu}m_{\alpha}m_{\beta}, \nonumber \\
\rho_{\alpha \beta \mu} &=& h_{ll}[n_{\alpha}n_{\beta}]_{;\mu} + 2 h_{lm}[n_{(\alpha}\bar{m}_{\beta)}]_{;\mu}+ 2 h_{l\bar{m}}[n_{(\alpha}m_{\beta)}]_{;\mu} \nonumber \\
			&+& h_{mm}[\bar{m}_{\alpha}\bar{m}_{\beta}]_{;\mu} +h_{\bar{m}\bar{m}}[m_{\alpha}m_{\beta}]_{;\mu}.
\end{eqnarray}
The first term appearing in Eq.~(\ref{eqn:Ttmp}) can be calculated exactly.  It is 
\begin{eqnarray} \label{eqn:eta2}
\eta^{\alpha \beta}_{\ \ \mu}\eta^{*}_{\alpha \beta \nu} &=& h_{mm,\mu}h^{*}_{\bar{m}\bar{m},\nu}+  h_{\bar{m}\bar{m},\mu}h^{*}_{mm,\nu}.
\end{eqnarray}
Evaluating the remaining terms requires more effort.  They involve products of tetrad vectors and their covariant derivatives. 
It is easy to show that the only non-vanishing $t$, $r$, 
and $\varphi$ components are $m^{\alpha}n_{\alpha;\varphi} = -(\sqrt{2}/4)\imath f\sin \theta$,  $\bar{m}^{\alpha}m_{\alpha;\varphi} = -\imath \cos\theta$, and 
$m^{\alpha}_{\ ;\varphi}n_{\alpha;r} = \sqrt{2}M/(4r^2)\imath \sin\theta$.
Using these and Eq.~(\ref{eqn:ORGrTetrad}), we find that 
\begin{eqnarray}
T_{tr}    &=& -\frac{1}{32\pi} (\dot{h}_{mm} \dot{h}^{*}_{\bar{m}\bar{m}}+ \dot{h}_{\bar{m}\bar{m}}\dot{h}^{*}_{mm}) \label{eqn:Ttrfz} \\
T_{r\varphi} &=&-\frac{1}{64\pi} (\dot{h}_{mm}h^{*}_{\bar{m}\bar{m},\varphi}+  \dot{h}_{\bar{m}\bar{m}}h^{*}_{mm,\varphi}) \nonumber \\
	       &+& \frac{\imath}{64\pi} ( \dot{h}_{mm}h^{*}_{\bar{m}\bar{m}}+  \dot{h}_{\bar{m}\bar{m}}h^{*}_{mm})\cos \theta +c.c. ,\label{eqn:Trphifz}
\end{eqnarray}
where we replaced $r$-derivatives with $t$-derivatives, and neglected terms of order ${\mathcal O}(r^{-3})$ and higher.

These expressions for the stress-energy tensor can be used to calculate the fluxes in the radiation zone.
Inserting Eq.~(\ref{eqn:Ttrfz}) into the first of Eq.~(\ref{eqn:FluxGeneral}), where we set $\epsilon=1$ and $f\approx 1$,
yields 
\begin{eqnarray} \label{eqn:EFluxFZ}
\frac{dE}{dt} &=& \frac{r^{2}}{32\pi}\int d\Omega \ (\dot{h}_{mm} \dot{h}^{*}_{\bar{m}\bar{m}}+ \dot{h}_{\bar{m}\bar{m}}\dot{h}^{*}_{mm}) \ = \ \frac{1}{32\pi}\int d\Omega \ \dot{h}_{AB}\dot{h}^{*AB} \nonumber \\
	      &=& \frac{1}{32\pi}\sum_{lm}\sum_{l'm'} \int d \Omega \ \Bigg [ |\dot{\psi}_{\rm ZM}|^2\ V^{lm}_{AB}V^{* AB}_{l'm'} +4 |\psi_{\rm RW}|^2 W^{lm}_{AB}W^{*AB}_{l'm'}\Bigg] \nonumber \\
	      &=& \frac{1}{64\pi}\sum_{lm}\frac{(l+2)!}{(l-2)!}\Bigg [ |\dot{\psi}_{\rm ZM}|^2+4 |\psi_{\rm RW}|^2\Bigg],
\end{eqnarray}
where in the first line we use  
$\Omega^{AC}\Omega^{BD}h_{AB}h^{*}_{CD}=r^2(h_{mm}h^{*}_{\bar{m}\bar{m}}+h_{\bar{m}\bar{m}}h^{*}_{mm})$, 
the second line follows from Eq.~(\ref{eqn:hORG}), 
and the third line follows from evaluating the angular 
integral with the aid of 
\[
\int d \Omega \ V^{lm}_{AB}\ V^{AB*}_{l'm'}\ =\ \frac{1}{2}\frac{(l+2)!}{(l-2)!}\delta_{ll'}\delta_{mm'}, \qquad \int d \Omega \ W^{lm}_{AB}\ W^{AB*}_{l'm'}\ =\ \frac{1}{2}\frac{(l+2)!}{(l-2)!}\delta_{ll'}\delta_{mm'}.
\]

The angular momentum flux calculation follows similar steps. 
Inserting Eq.~(\ref{eqn:Trphifz}) into the second of Eq.~(\ref{eqn:FluxGeneral}), we get 
\begin{eqnarray} \label{dLtmp}
\frac{dL}{dt} &=& -\frac{r^2}{64\pi}\int d\Omega \ \Bigg[(\dot{h}_{mm}h^{*}_{\bar{m}\bar{m},\varphi}+  \dot{h}_{\bar{m}\bar{m}}h^{*}_{mm,\varphi}) \nonumber \\
	      && \qquad -\imath ( \dot{h}_{mm}h^{*}_{\bar{m}\bar{m}}+  \dot{h}_{\bar{m}\bar{m}}h^{*}_{mm})\cos \theta\Bigg]+c.c.
\end{eqnarray}
The last term involves the product of $\cos\theta$ with a term of the form $\sin\theta S^{lm}(\theta)S^{*}_{l'm'}(\theta)$, where $S^{lm}(\theta)$ 
is a spherical harmonic function.  Under the interchange $\theta\rightarrow \pi-\theta$, 
we have that $\cos\theta \rightarrow -\cos\theta$ and $\sin\theta S^{lm}(\theta)S^{*}_{l'm'}(\theta)\rightarrow \sin\theta S^{lm}(\theta)S^{*}_{l'm'}(\theta)$.  
The overall term is therefore odd in $\theta$ with respect 
to $\pi/2$, and  integration over $0\leq \theta \leq \pi$ yields zero contribution to the angular momentum flux.
We are then left with 
\begin{eqnarray} \label{eqn:LFluxFZ}
\frac{dL}{dt} &=& -\frac{r^2}{64\pi}\int d\Omega \ (\dot{h}_{mm}h^{*}_{\bar{m}\bar{m},\varphi}+  \dot{h}_{\bar{m}\bar{m}}h^{*}_{mm,\varphi}+c.c.) \nonumber \\
	      &=& \frac{\imath m r^2}{64\pi}\int d\Omega \ (\dot{h}_{mm}h^{*}_{\bar{m}\bar{m}}+  \dot{h}_{\bar{m}\bar{m}}h^{*}_{mm})+c.c.) \nonumber \\
	      &=& \frac{\imath m}{64\pi} \int d\Omega \ (\dot{h}^{AB}h^{*}_{AB} +c.c.) \nonumber \\
	      &=&\sum_{lm}\sum_{l'm'} \frac{\imath m}{64\pi} \int d\Omega \  \Bigg[ \dot{\psi}_{\rm ZM}\psi^{*}_{\rm ZM}\ V^{lm}_{AB}V^{* AB}_{l'm'}+4 \psi_{\rm RW} \int_{-\infty}^{t} dt' \psi^{*}_{\rm RW}(t') W^{AB}_{lm}W^{*AB}_{l'm'}\Bigg] +c.c. \nonumber \\
	      &=&\sum_{lm} \frac{\imath m }{128 \pi} \frac{(l+2)!}{(l-2)!} \Bigg[\dot{\psi}_{\rm ZM}\ \psi^{*}_{\rm ZM} +4 \psi_{\rm RW}\  \int_{-\infty}^{t} dt' \psi^{*}_{\rm RW}(t')\Bigg] +c.c.\ ,
\end{eqnarray}
where we use $h_{\bar{m}\bar{m},\varphi}=-\imath m h_{\bar{m}\bar{m}}$ in the first line, and the 
remaining steps shadow the ones for the energy flux calculation.

%%%%%%%%%%%%%%%%%%%%%%%%%%%%%%%%%%%%%%%%%%%%%%%%%%%%%%%%%%%%%%%%%%%%%%%%%%%%
%%%%%%%%%%%%%%%%%%%%%%%%%%%%%%%%%%%%%%%%%%%%%%%%%%%%%%%%%%%%%%%%%%%%%%%%%%%%

\subsection{Black hole absorption} \label{app:irg}
The calculation of the black hole absorption is similar to the calculation 
of the far-zone fluxes.  Here we are looking to isolate the divergent 
piece of $T_{\mu\nu}$, since it is this part that corresponds to the blueshifted gravitational waves: 
The expansion parameter is $f$, and we are looking for the ${\mathcal O}(f^{-1})$ portion of 
the $tr$ and $r\varphi$ components of $T_{\mu\nu}$.  
We neglect terms of order ${\mathcal O}(1)$.

The material developed in  Sec.~\ref{app:org} 
can be used here simply by 
replacing $l^{\mu} \leftrightarrow n^{\mu}$.
In an ingoing radiation gauge, 
the non-trivial tetrad components of the perturbation tensor are
\begin{eqnarray}\label{eqn:IRGrtetrad}
h_{nn} &\sim& {\mathcal O}(f), \nonumber \\
h_{nm} &\sim& {\mathcal O}(f), \nonumber \\
h_{\bar{m}\bar{m}} &=& \frac{1}{4M^2}h^{IRG}_{AB}\ \bar{m}^{A}\bar{m}^{B},\nonumber \\
h_{mm} &=& \frac{1}{4M^2}h^{IRG}_{AB}\ m^{A}m^{B},
\end{eqnarray}
and $m^{A}$ was introduced previously.

With the replacement 
$n^{\nu} \leftrightarrow l^{\nu}$, the steps we follow are almost exactly the same as those of the radiation zone calculations.  
The stress-energy tensor is written as in Eq.~(\ref{eqn:Ttmp}), with $\eta_{\alpha \beta\mu}$ and $\rho_{\alpha \beta \nu}$ 
changed to reflect the exchange of tetrad vectors.  
It is not difficult to show that the non-vanishing components of the contracted derivatives 
of the tetrad vectors are now $m^{\alpha}l_{\alpha;\varphi} = (\sqrt{2}/2)\imath\sin\theta$, 
$\bar{m}^{\alpha}m_{\alpha;\varphi} = -\imath \cos\theta$, and $ m^{\alpha}_{\ ;\varphi}l_{\alpha;r} = \sqrt{2}M/(2r^2)\imath f^{-1}\sin\theta$.

These, combined with Eq.~(\ref{eqn:IRGrtetrad}), reveal that the relevant components of the stress-energy tensor 
are given by 
\begin{eqnarray} \label{eqn:Tblue}
T_{tr}    &=& -\frac{f^{-1}}{32\pi} (\dot{h}_{mm} \dot{h}^{*}_{\bar{m}\bar{m}}+ \dot{h}_{\bar{m}\bar{m}}\dot{h}^{*}_{mm}) \label{eqn:Ttreh} \\
T_{r\varphi} &=& -\frac{f^{-1}}{64\pi} (\dot{h}_{mm}h^{*}_{\bar{m}\bar{m},\varphi}+  \dot{h}_{\bar{m}\bar{m}}h^{*}_{mm,\varphi}) \nonumber \\
	       &+& \frac{\imath f^{-1}}{64\pi} ( \dot{h}_{mm}h^{*}_{\bar{m}\bar{m}}+  \dot{h}_{\bar{m}\bar{m}}h^{*}_{mm})\cos \theta +c.c. \, , \label{eqn:Trphieh}
\end{eqnarray}
where we replaced $r^{*}$-derivatives with $t$-derivatives and neglected term of order ${\mathcal O}(1)$.
Note that this is exactly of the same form (apart from a factor of $f^{-1}$)  
as that obtained for $T_{tr}$ and $T_{r\varphi}$ in the far zone. 

To calculate the fluxes, we insert these expressions into 
Eq.~(\ref{eqn:FluxGeneral}), where we also 
set $\epsilon =-1$. The divergence in the stress-energy tensor is 
canceled by the factor of $f$ appearing in Eq.~(\ref{eqn:FluxGeneral}).  
The remaining calculations are identical with those of the far-zone, 
with $h^{IRG}_{AB}$ given by Eq.~(\ref{eqn:hIRG}).  The energy flux is 
\begin{eqnarray} \label{eqn:EFluxEH}
\frac{dE}{dt} &=& \sum_{lm}\frac{1}{64\pi}\frac{(l+2)!}{(l-2)!}\Bigg [ |\dot{\psi}_{\rm ZM}|^2+4 |\psi_{\rm RW}|^2\Bigg],
\end{eqnarray}
while the angular momentum flux is 
\begin{eqnarray} \label{eqn:LFluxEH}
\frac{dL}{dt} &=& \sum_{lm}\frac{\imath m }{128 \pi} \frac{(l+2)!}{(l-2)!} \Bigg[\dot{\psi}_{\rm ZM}\ \psi^{*}_{\rm ZM} +4 \psi_{\rm RW}\  \int_{-\infty}^{t} dt' \psi^{*}_{\rm RW}(t')\Bigg] +c.c.\ .
\end{eqnarray}

\end{document}